\documentclass[11pt]{article}
\DeclareMathAlphabet{\scr}{U}{rsfs}{m}{n}
\usepackage{latexsym}
\usepackage[mathscr]{eucal}
\usepackage{amsfonts}
\usepackage{amscd}
\usepackage{cite}
\usepackage{array}   
\usepackage{amssymb}
\usepackage{colordvi}
\usepackage[centertags]{amsmath}
\usepackage{enumerate}
\usepackage{graphicx}
\usepackage{booktabs}
\usepackage{theorem}
\usepackage[footnotesize]{caption}
\usepackage{soul}

\newcommand{\cleqn}{\setcounter{equation}{0}}
\newcommand{\newc}{\newcommand}
\newc{\be}{\begin{equation}}
\newc{\ee}{\end{equation}}
\newc{\bea}{\begin{eqnarray}}
\newc{\eea}{\end{eqnarray}}
\newc{\ol}{\overline}
\newc{\wt}{\widetilde}
\newc{\bs}{\boldsymbol}
\newc{\m}{\mathcal}
\newc{\la}{\langle}
\newc{\ra}{\rangle}

\newcommand{\lsim}{\raisebox{-0.13cm}{~\shortstack{$<$ \\[-0.07cm] $\sim$}}~} 
\newcommand{\gsim}{\raisebox{-0.13cm}{~\shortstack{$>$ \\[-0.07cm] $\sim$}}~}

\newcommand{\non}{\nonumber} 
\newcommand{\beq}{\begin{eqnarray}} 
\newcommand{\eeq}{\end{eqnarray}}

\newcommand{\sn}{\newline \vspace*{-3.5mm}}

\newcommand{\bc}{\begin{center}}
\newcommand{\ec}{\end{center}}

\newcommand{\ba}{\begin{array}}
\newcommand{\ea}{\end{array}}
\newcommand{\ds}{\displaystyle}
\begin{document}

\title{\hfill ~\\[-30mm]
% \hfill\mbox{\small SHEP-11-36}\\[-3.5mm]
%\hfill\mbox{\small IPPP-11-83}\\[-3.5mm]
\phantom{h} \hfill\mbox{\small KA-TP-01-2012}\\[-1.1cm]
\phantom{h} \hfill\mbox{\small SFB/CPP-12-02} \\[-1.1cm]
\phantom{h} \hfill\mbox{\small UH511-1188-2012}
\\[1cm]
%\title{
\vspace{13mm}   \textbf{NMSSM Higgs Benchmarks Near 125 GeV \\[4mm]}}

\date{}
\author{
S.~F.~King$^{1\,}$\footnote{E-mail: \texttt{king@soton.ac.uk}},
M. M\"{u}hlleitner$^{2\,}$\footnote{E-mail: \texttt{maggie@particle.uni-karlsruhe.de}},
R.~Nevzorov$^{3\,}$\footnote{E-mail: \texttt{nevzorov@phys.hawaii.edu}}
\footnote{On leave of absence from the Theory Department,
ITEP, Moscow, Russia.
}
\\[9mm]
{\small\it
$^1$School of Physics and Astronomy,
University of Southampton,}\\
{\small\it Southampton, SO17 1BJ, U.K.}\\[3mm]
{\small\it
$^2$Institute for Theoretical Physics, Karlsruhe Institute of Technology,} \\
{\small\it 76128 Karlsruhe, Germany.}\\[3mm]
{\small\it
$^3$Department of Physics and Astronomy, University of Hawaii,}\\
{\small\it Honolulu, HI 96822, Hawaii, United States.}\\
}

\maketitle

\begin{abstract}
\noindent  
The recent LHC indications of a SM-like Higgs boson near 125 GeV are
consistent not only with the Standard Model (SM) but also with Supersymmetry (SUSY).
However naturalness arguments disfavour the Minimal Supersymmetric Standard Model (MSSM).
We consider the Next-to-Minimal Supersymmetric Standard Model (NMSSM)
with a SM-like Higgs boson near 125 GeV involving relatively light
stops and gluinos below 1 TeV in order to satisfy naturalness requirements. 
We are careful to ensure that the chosen values of couplings do not become non-perturbative
below the grand unification (GUT) scale, although we also examine how
these limits may be extended by the addition 
of extra matter to the NMSSM at the two-loop level. We then propose
four sets of benchmark points corresponding to the SM-like Higgs boson
being the lightest or the second lightest Higgs state in the NMSSM or
the NMSSM-with-extra-matter. With the aid of these benchmark points we 
discuss how the NMSSM Higgs boson near 125 GeV may be distinguished
from the SM Higgs boson in future LHC searches.
% by measuring its cross-section and branching ratios.  
%For example, the Higgs branching ratio into two photons may easily be enhanced by a factor of two in the NMSSM.
 \end{abstract}
\thispagestyle{empty}
\vfill
\newpage
\setcounter{page}{1}

%%%%%%%%%%%%%%%%%%%%%%%%%%%%%%%%%%%%%%%%%%%%%%%%%%%%%%%%%%%%%%%%%%%
\section{Introduction}
The ATLAS and CMS Collaborations have recently presented the first
indication for a Higgs boson with a mass in the region
$\sim124-126$~GeV \cite{AtlasTalk,CMSTalk}. An excess of events is
observed by the ATLAS experiment for a Higgs boson mass hypothesis
close to 126 GeV with a maximum local statistical significance of 3.6$\sigma$
above the expected SM background and by the CMS experiment at 124 GeV with
2.6$\sigma$ maximum local significance. If the ATLAS and CMS signals
are combined the statistical significance increases, but is still 
less than the 5$\sigma$ required to claim a discovery. Interestingly, 
the ATLAS signal in the $\gamma \gamma$ decay channel by itself has a
local significance of 2.8$\sigma$ whereas a SM-like Higgs boson would
only have a significance of half this value, leading to speculation
that the observed Higgs boson is arising from beyond SM physics. 
In general, these results have generated much excitement in the
community, and already there are a number of papers discussing the
implications of such a Higgs boson
\cite{Higgs,Hall:2011aa,Arvanitaki:2011ck,Ellwanger:2011aa,Gunion:2012zd}.
%A well known observation is that while the LHC can in principle
%exclude a Standard Model (SM) Higgs boson, it can only discover a
%SM-like Higgs boson, which could for example correspond to a SUSY
%Higgs boson near the decoupling region\cite{Haber}. 
\sn

In the Minimal Supersymmetric Standard Model (MSSM) the lightest Higgs
boson is lighter than about 130-135 GeV, depending on top squark
parameters (see e.g.~\cite{Djouadi:2005gj} and references therein).
A 125 GeV SM-like Higgs boson is consistent with the MSSM in the
decoupling limit. In the limit of decoupling the light Higgs mass is given by
\begin{equation}
m_h^2 \, \approx  \, M_Z^2 \cos^2 2 \beta + \Delta m_h^2 \; , 
\label{eq:hmassMSSM}
\end{equation}
where $ \Delta m_h^2$ is dominated by loops of heavy top quarks and
top squarks and $\tan \beta$ is the ratio of the vacuum expectation
values (VEVs) of the two Higgs doublets introduced in the MSSM Higgs
sector. At large $\tan \beta$, we require $\Delta m_h \approx 85$~GeV
which means that a very substantial loop contribution, nearly as large
as the tree-level mass, is needed to raise the Higgs boson mass to 125
GeV.  
%Due to the logarithmic dependence of the Higgs boson mass on the
%stop masses the latter depend exponentially on the Higgs boson mass. 
The rather complicated parameter dependence has been studied in
\cite{Hall:2011aa} where it was shown that, with ``maximal stop
mixing'',   the lightest stop mass must be $m_{\tilde t_1}\gsim 500$
GeV (with the second stop mass considerably larger) in the MSSM in order to achieve
a 125 GeV Higgs boson. However one of the motivations for SUSY is to
solve the hierarchy or fine-tuning problem of the SM
\cite{CERN-TH-4825/87}. It is well known that such large stop masses
typically require a tuning at least of order 1\% in the MSSM,
depending on the parameter choice and the definition of fine-tuning
\cite{SusyFineTuning}. \sn

In the light of such fine-tuning considerations, it has been known for some time, even after the 
LEP limit on the Higgs boson mass of 114 GeV, that the fine-tuning of the
MSSM could be ameliorated in the Next-to-Minimal Supersymmetric
Standard Model (NMSSM) \cite{NMSSMtuning}. With a 125 GeV Higgs boson,
this conclusion is greatly strengthened and the NMSSM appears to be a
much more natural alternative. In the NMSSM, the spectrum of the MSSM is extended by
one singlet superfield~\cite{genNMSSM1,genNMSSM2,Nevzorov:2004ge} (for reviews see
\cite{Maniatis:2009re ,Ellwanger:2009dp}). In the NMSSM the
supersymmetric Higgs mass parameter $\mu$ is promoted to a
gauge-singlet superfield, $S$, with a coupling to the Higgs doublets,
$\lambda S H_u H_d$, that is perturbative up to unified scales.
In the pure NMSSM values of
$\lambda \sim 0.7$ do not spoil the validity of perturbation 
theory up to the GUT scale only providing $\tan\beta\gtrsim 4$,
however the presence of additional extra matter \cite{Barbieri:2007tu} allows smaller values of $\tan\beta$
to be achieved.
%In this case the 125 GeV Higgs boson can be present in the particle spectrum 
%even when the mixing in the stop sector is relatively small. 
%The phenomenological implications of the NMSSM with extra exotic matter
%were considered in. \sn
%thereby constraining $\lambda \lesssim 0.7$ (everywhere in this paper
%$\lambda$ refers to the weak scale value of the coupling).  
The maximum mass of the lightest Higgs boson is  
\begin{equation}
m_h^2 \, \approx  \,  M_Z^2 \cos^2 2 \beta + \lambda^2 v^2 \sin^2 2 \beta +  \Delta m_h^2 \;
\label{eq:hmassNMSSM}
\end{equation}
where here we use $v=174$~GeV.  For $\lambda v > M_Z$, the tree-level
contributions to $m_h$ are maximized for moderate values of $\tan \beta$ 
rather than by large values of $\tan \beta$ as in the MSSM. For example, taking
$\lambda =0.7$ and $\tan\beta=2$, these tree-level
contributions raise the Higgs boson mass to about 112 GeV, 
and $\Delta m_h \gtrsim 55\,\mbox{GeV}$ is required. This is to be compared 
to the MSSM requirement $\Delta m_h \gtrsim 85\,\mbox{GeV}$. The difference 
between these two values (numerically about 30 GeV) is significant since
$\Delta m_h$ depends logarithmically on the stop
masses as well as receiving an important contribution from stop
mixing. This means for example, that, unlike the MSSM, in the case of
the NMSSM maximal stop mixing is not required to get the Higgs heavy
enough. \sn

The NMSSM has in fact several attractive features as
compared  to the widely studied MSSM. Firstly, the NMSSM naturally solves in an
elegant way the so--called $\mu$ problem~\cite{MuProblem} of the MSSM, namely that 
the phenomenologically required value for the supersymmetric Higgsino mass $\mu$ 
in the vicinity of the electroweak or SUSY breaking scale is not explained. 
This is automatically achieved in the NMSSM, since an effective $\mu$-parameter is
dynamically generated when the singlet Higgs field acquires a vacuum expectation
value of the order of the SUSY breaking scale, leading to a
fundamental Lagrangian that  contains no dimensionful parameters apart
from the soft  SUSY breaking terms. Secondly, as compared to the MSSM,
the NMSSM can induce a richer phenomenology in the Higgs and
neutralino sectors, both in collider and dark 
matter (DM) experiments. For example, heavier Higgs states can decay into
lighter ones with sizable rates~\cite{lighthiggs1,NoLoseNMSSM2,
nmhdecay1,FineTuning1,egh1,lighthiggs2,lighthiggsteva1,lighthiggs3,lighthiggsteva2,lighthiggs4,lighthiggsteva3,cheung}. 
In addition, a new possibility appears for achieving the correct
cosmological relic density~\cite{relicdens} through the 
so-called ``singlino", i.e. the fifth neutralino of the model, which can have
weaker-than-usual couplings to SM particles.
The NMSSM Higgs boson near 125 GeV may also have significantly different properties than
the SM Higgs boson \cite{Arvanitaki:2011ck,Ellwanger:2011aa}, as we
shall discuss in some detail here. Thirdly, as already discussed
at length above, the NMSSM requires less fine-tuning than the MSSM \cite{NMSSMtuning}.
\sn

Since the NMSSM (as in the case for the MSSM) has a rich parameter space,
it is convenient to consider the standard approach of so-called ``benchmark
points'' in the SUSY parameter space. These consist of a few ``discrete"  parameter
configurations of a  given SUSY model, which in our case are supposed to lead to 
typical phenomenological features. Using discrete points avoids
scanning the entire parameter space, focusing instead on
representative choices that reflect the new interesting  
features of the model, such as new signals, peculiar mass spectra, and so on. A
reduced number of points can then be subject to full experimental investigation,
without loss of substantial theoretical information.
While several such benchmark scenarios have been devised for the 
MSSM~\cite{SUSYbenchmarks,MSSMBenchmarks}  and thoroughly studied in both the
collider  and the DM contexts, and even for the NMSSM \cite{Djouadi:2008uw},
these need to be revised in the light of the 125 GeV Higgs signal.
The motivation for finding such NMSSM benchmarks is to enable the
characteristics of the NMSSM Higgs boson near 125 GeV to be identified, so
that it may eventually be resolved from that of the SM Higgs boson.  
\sn

In this paper we present a set of NMSSM benchmark points with a
SM-like Higgs boson mass near 125 GeV, focussing on the cases where
both stop masses are as light as possible in order to reduce the
fine-tuning, in accordance with the above discussion. The tools to
calculate the Higgs and SUSY particle spectra in the NMSSM, in
particular {\tt NMSSMTools}, have been 
available for some time~\cite{nmhdecay1,nmhdecay2,nmssmtools},
although these will be supplemented by other codes as discussed later.
The goal of the paper is to firstly find NMSSM benchmark points
which contain a 125 GeV SM-like Higgs boson 
and in addition involve relatively light stops. Since the stops
receive radiative corrections to their masses from gluinos, we shall
also require relatively light gluinos as well as having as small an 
effective $\mu_{\rm eff}$ parameter ($\mu_\mathrm{eff} = \lambda \langle S \rangle $) 
as possible \cite{Brust:2011tb}. Having relatively light stops (and
sbottoms), care has to be taken not to be in conflict with direct searches at the Tevatron
and recent searches at the LHC. In addition, light stops appearing in
the triangle loop diagrams can significantly affect Higgs production 
via gluon fusion as well as Higgs decay into two photons. 
Once a parameter set leading to a 125 GeV NMSSM Higgs boson has been
found it has to be checked if the production cross-sections and
branching ratios are such that they lead to a total Higgs production
cross-section times branching ratios which is compatible with the LHC
searches. The obtained branching ratios can be used to ultimately
distinguish the NMSSM Higgs boson from the SM Higgs state. However, we do
not attempt to simulate the LHC search strategies, our goal being the
much more modest one of providing benchmarks with different
characteristic types of NMSSM Higgs bosons, which can eventually be studied
in more detail. The types of benchmark points considered here all
involve relatively large values of $\lambda > 0.5$ and small values 
of $\tan \beta \le 3$, in order to allow for the least fine-tuning
possible involving light stops as discussed above. However we are
careful to respect the perturbativity bounds on $\lambda$ up to the
GUT scale, either for the pure NMSSM, or the NMSSM supplemented by
three copies of extra $SU(5)$ $5+\overline{5}$ states, where such
bounds are calculated using two-loop renormalisation group equations
(RGEs). We find that the NMSSM Higgs boson near 125 GeV can come in
many guises. It may be very SM-like, practically indistinguishable
from the SM Higgs boson. Or it may have different Higgs production
cross-sections and widely varying decay branching ratios which enable it to be 
resolved from the SM Higgs boson. The key distinguishing feature of
the NMSSM, compared to the MSSM, is the presence of the singlet $S$
which may mix into the 125 GeV Higgs mass state to a greater or lesser extent. 
As the singlet component of the 125 GeV Higgs boson is increased,
either the $H_d$ or $H_u$ components (or both) must be reduced. If the
$H_d$ component is reduced then this reduces the decay rate into bottom
quarks and can allow other rare decays like $\gamma \gamma$ to have
larger branching ratios  \cite{Ellwanger:2011aa}.  
In addition there is the effect of SUSY particle and charged Higgs
boson loops in the Higgs coupling to photons which can increase or
decrease the rate of decays into $\gamma \gamma$. \sn

In order to put the present work in context, it is worth to emphasise
that this is the first paper to propose ``natural'' (i.e. involving
light stops) NMSSM benchmark points with a SM-like Higgs boson near
125 GeV.  For example, the previous NMSSM benchmark paper
\cite{Djouadi:2008uw} was concerned with the constrained NMSSM which
does not allow light stops consistent with current LHC SUSY
limits. Also, while this paper was being prepared there have appeared
a number of other related papers, none of which however are focussed
on the task of providing benchmark points. For example, the results in
\cite{Hall:2011aa} were mainly concerned with fine-tuning issues in
SUSY models with a $\lambda SH_uH_d$ coupling for a SM-like Higgs
boson near 126 GeV and the actual NMSSM was strictly not considered
since an explicit $\mu$ term was included whereas the trilinear
singlet coupling $\frac{\kappa}{3} S^3$ was neglected. 
By contrast another recent paper \cite{Arvanitaki:2011ck} did consider
the actual NMSSM although no benchmarks were proposed.
As this paper was being finalised, further dedicated NMSSM papers with
a Higgs boson near 125 GeV have started to appear, in particular
 \cite{Ellwanger:2011aa} in which the two photon enhancement was
 emphasised but without benchmark points, and \cite{Gunion:2012zd}
 where (versions of) the constrained NMSSM were considered. It should
 be clear that the present paper is both complementary and
 contemporary to all these papers.  \sn

The layout of the remainder of the paper is as follows. In Section~\ref{mssm} we discuss the MSSM and fine-tuning, showing that it leads to constraints on the mass of the heavier stop quark. 
In Section~\ref{nmssm} we briefly review the NMSSM, and provide
perturbativity limits of the coupling $\lambda$ depending on the
coupling $\kappa$ as well as $\tan \beta$. We show how the limit on $\lambda$ 
may be increased if there is extra matter in the SUSY desert. 
Section~\ref{sec:constraints} is concerned with constraints from Higgs
searches, SUSY particle searches and Dark Matter. Section~\ref{bm}
contains the four sets of NMSSM Higgs benchmarks near 125 GeV that we
are proposing, including the Higgs production cross-sections and
branching ratios which ultimately will enable it to be distinguished
from the SM Higgs boson. Section~\ref{summary} summarises and concludes the paper. 
Appendix~\ref{sec} contains the two-loop RGEs from which the
perturbativity bounds on $\lambda$ were obtained. 
 
%%%%%%%%%%%%%%%%%%%%%%%%%%%%%%%%%%%%%%%%%%%%%%%%%%%%%%%%%%%%%%%%%%%
\section{The MSSM \label{mssm}}
The superpotential of the MSSM is given, in terms of (hatted)
superfields, by
\beq
{\cal W} = \mu \widehat{H}_u \widehat{H}_d + 
h_t \widehat{Q}_3\widehat{H}_u\widehat{t}_R^c - h_b \widehat{Q}_3
\widehat{H}_d\widehat{b}_R^c  - h_\tau \widehat{L}_3 \widehat{H}_d
\widehat{\tau}_R^c \; , 
\label{supot1}
\eeq
in which only the third generation fermions have been included
(with possible neutrino Yukawa couplings have been set to zero), and $\widehat
Q _3,  \widehat L _3$ stand for superfields associated with the $(t,b)$ and
$(\tau,\nu_\tau)$  SU(2) doublets. \sn

The soft SUSY breaking terms consist of  the scalar mass terms for the Higgs and sfermion
scalar fields which, in terms of the  fields corresponding to the complex scalar
components of the superfields, are given by
\beq
 -{\cal L}_\mathrm{mass} &=& 
 m_{H_u}^2 | H_u |^2 + m_{H_d}^2 | H_d|^2  \non \\
 &+& m_{{\tilde Q}_3}^2|{\tilde Q}_3^2| + m_{\tilde t_R}^2 |{\tilde t}_R^2|
 +  m_{\tilde b_R}^2|{\tilde b}_R^2| +m_{{\tilde L}_3}^2|{\tilde L}_3^2| +
 m_{\tilde  \tau_R}^2|{\tilde \tau}_R^2|\; ,
\eeq
and the trilinear interactions between the sfermion and Higgs fields, 
\beq
-{\cal L}_\mathrm{tril}=  B\mu H_u H_d  + h_t A_t \tilde Q_3 H_u \tilde t_R^c - h_b A_b
\tilde Q_3 H_d \tilde b_R^c - h_\tau A_\tau \tilde L_3 H_d \tilde \tau_R^c
+ \mathrm{h.c.}\;.
\eeq
The tree-level MSSM Higgs potential is given by
\be
V_0=m_1^2|H_d|^2+m_2^2|H_u|^2-m_3^2(H_d H_u+h.c.)+ 
\frac{g_2^2}{8}\left(H_d^+\sigma_a 
H_d+H_u^+\sigma_a
H_u\right)^2
+\frac{{g'}^2}{8}\left(|H_d|^2-|H_u|^2\right)^2
\label{7}
\ee
where $g'=\sqrt{3/5} g_1$, $g_2$ and $g_1$ are the low energy (GUT normalised)
$SU(2)_W$ and $U(1)_Y$ gauge couplings, $m_1^2=m_{H_d}^2+\mu^2$,
$m_2^2=m_{H_u}^2+\mu^2$ and $m_3^2=-B\mu$. \sn

In the MSSM, at the 1-loop level, stops contribute to the Higgs boson
mass and three more parameters become important, the stop soft masses,
$ m_{{\tilde Q}_3}$ and $m_{\tilde t_R}$, and the stop mixing parameter 
\be
\label{Xt}
X_t = A_t - \mu \cot \beta .
\ee
The dominant one-loop contribution to the Higgs boson mass depends on
the geometric mean of the stop masses, $m_{\tilde t}^2 =
m_{\tilde{Q}_{3}} m_{\tilde{t}_R}$, and is given by, 
\be \label{eq:HiggsMassCorr}
\Delta m_h^2 \approx \frac{3}{(4 \pi)^2} \frac{m_t^4}{v^2} \left[ \ln \frac{m_{\tilde t}^2}{m_t^2} + \frac{X_t^2}{m_{\tilde t}^2} \left(1-\frac{X_t^2}{12 m_{\tilde t}^2} \right)\right].
\ee
The Higgs mass is sensitive to the degree of stop mixing through the second term in the brackets, and is maximized for $|X_t| = X_t^{\rm max} =  \sqrt 6 \, m_{\tilde t}$, which was referred to as ``maximal mixing" above.  \sn
 
\begin{figure}
\bc
\hspace*{-11cm}{$\Delta_{II}$}\\
\includegraphics[height=70mm,keepaspectratio=true]{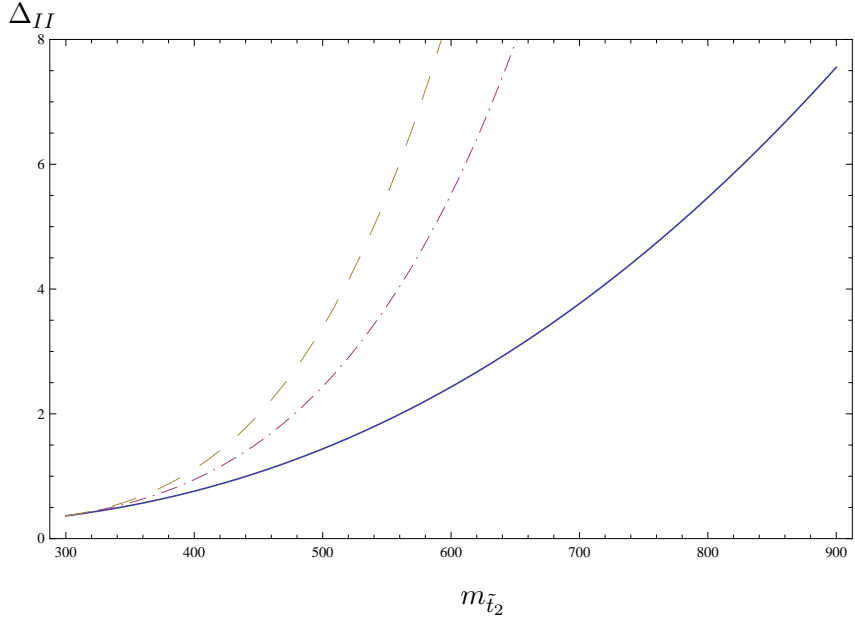}\\[1mm]
\hspace*{1cm}{$m_{\tilde{t}_2}$}\\[1mm]
\caption{\label{ft2} The contribution of one--loop corrections to Eq.~(\ref{361})
as a function of the mass of the heaviest stop $m_{\tilde{t}_2}$ for $\tan\beta=10$,
$\mu=200\,\mbox{GeV}$, $Q=m_t=165\,\mbox{GeV}$ 
and $m_{\tilde{t}_1}=300\,\mbox{GeV}$. Here $\Delta_{II}=2\cdot \Delta/M_Z^2$. Solid, 
dashed--dotted and dashed lines correspond to the mixing angle in the stop sector 
$\theta_t=0,\pi/8$ and $\pi/4$, respectively.}
\ec
\end{figure}

The fine-tuning in the MSSM can be simply understood by examining the  
leading one--loop correction to the Higgs potential, 
\begin{eqnarray}
\Delta V=\frac{3}{32\pi^2}\left[m_{\tilde{t}_1}^4\left(
\ln\frac{m_{\tilde{t}_1}^2}{Q^2}-\frac{3}{2}\right)+
m_{\tilde{t}_2}^4\left(\ln\frac{m_{\tilde{t}_2}^2}{Q^2}-\frac{3}{2}\right)
-2m_t^4\left(\ln\frac{m_t^2}{Q^2}-\frac{3}{2}\right)\right]\,,
\label{14}
\end{eqnarray}
where the two stop masses are,
\begin{eqnarray}
\hspace{-20mm}
m_{\tilde{t}_{1,2}}^2&=&\frac{1}{2}\left(m_{{\tilde Q}_3}^2+ m_{\tilde t_R}^2+2m_t^2 +\frac{1}{2}M_Z^2\cos 2\beta 
\right.
\nonumber\\
&&\left. 
\mp\sqrt{\left(m_{{\tilde Q}_3}^2 - m_{\tilde t_R}^2 +\frac{4}{3}M_W^2\cos 2\beta - \frac{5}{6}M_Z^2\cos 
2\beta\right)^2 + 4m_t^2 \, X_t^2}\right)\,.
\label{13}
\end{eqnarray}
The fine-tuning originates from the fact that $\Delta V \gg v^4$, where $v= 174$ GeV is the 
combined Higgs VEV. By considering the minimization conditions for the Higgs potential,
one finds
\begin{eqnarray}
\mu^2 + \frac{1}{2}M_Z^2+\Delta=\frac{(m_{H_d}^2-m_{H_u}^2\tan^2\beta)}{\tan^2\beta-1}\,,
\label{361}
\end{eqnarray}
where
\begin{eqnarray}                    
\Delta&=&\frac{1}{\tan^2\beta - 1}\left\{\frac{3 m_t^2}{16\pi^2 v^2 \cos^2\beta}\Biggl[f(m_{\tilde{t}_{2}})+
f(m_{\tilde{t}_{1}})-2f(m_t)\Biggr]\right.\nonumber\\
&&\left.
-\frac{3 M_Z^2}{64\pi^2 v^2 \cos^2\beta}\Biggl[f(m_{\tilde{t}_{2}})+
f(m_{\tilde{t}_{1}})\Biggr]\right.\nonumber\\
&&\left.-\frac{3}{32\pi^2 v^2 \cos^2\beta}\left(\frac{4}{3}M_W^2-\frac{5}{6}M_Z^2\right)\Biggl[f(m_{\tilde{t}_{2}})
-f(m_{\tilde{t}_{1}})\Biggr]\cos 2\theta_t\right.\nonumber\\
&&\left.
+\left(\frac{3(m_{\tilde{t}_{2}}^2-m_{\tilde{t}_{1}}^2) }{64\pi^2 v^2 \cos^2 \beta} \sin^2 2\theta_t+
\frac{3\, m_t\, \mu\, \sin 2\theta_t}{8\pi^2 v^2 \sin 2\beta}
\right)\Biggl[f(m_{\tilde{t}_{2}})-f(m_{\tilde{t}_{1}})\Biggr]
\right\}\,.
\label{362}
\end{eqnarray}
In Eq.~(\ref{362}) $\theta_t$ is the mixing angle in the stop sector given by
\begin{eqnarray}
\sin 2\theta_t = \frac{2 m_t X_t}{(m_{\tilde{t}_{2}}^2-m_{\tilde{t}_{1}}^2)}\,,
\label{363}   
\end{eqnarray} 
whereas
$$
f(m)= m^2 \left(\ln\frac{m^2}{Q^2}-1\right)\,.
$$
Here we set the renormalisation scale $Q=m_t$. From Eq.~(\ref{361})
one can see that in order to avoid tuning, 
\begin{eqnarray}
\Delta\lsim \frac{1}{2}M_Z^2.
\end{eqnarray}
%For example, by examining the first derivative of the potential correction,
%one finds, for large $\tan \beta$, setting the renormalisation scale $Q=m_t$, 
%the following constraint in order to avoid tuning,
%\begin{eqnarray}
%\Delta \equiv \frac{3 h_t^2}{16\pi^2}\left[m_{\tilde{t}_1}^2\left(
%\ln\frac{m_{\tilde{t}_1}^2}{m_t^2}-1\right)+
%m_{\tilde{t}_2}^2\left(\ln\frac{m_{\tilde{t}_2}^2}{m_t^2}-1\right)
%+2m_t^2\right] \lsim \frac{1}{2}M_Z^2.
%\label{36}
%\end{eqnarray}
This shows that both stop masses must be light to avoid tuning. 
For example, defining $\Delta_{II}=2\cdot \Delta/M_Z^2$ the absence 
of any tuning requires $\Delta_{II}\lsim 1$. This in turn requires 
the heavier stop mass to be below about 500 GeV as illustrated in 
Fig.\ref{ft2}. This constraint on the heavier stop mass has not been 
emphasised in the literature, where often the focus of attention is 
on the lightest stop mass. \sn

It has been noted that large or maximal stop mixing is associated 
with large fine-tuning. This also follows from Fig.\ref{ft2} and
Eq.~(\ref{362}). Indeed, Fig.\ref{ft2} demonstrates that the 
contribution of one--loop corrections to Eq.~(\ref{361}) increases
when the mixing angle in the stop sector becomes larger.
In fact when $\theta_t$ is close to $\pi/4$ the last term in 
Eq.~(\ref{362}) gives the dominant contribution to $\Delta$
enhancing the overall contribution of loop corrections 
in the minimization condition (\ref{361}) which determines the 
mass of the $Z$--boson. \sn
%, and 
%this is the reason, namely that the heavier stop mass becomes 
%even larger in the presence of large stop mixing (see Eq.\ref{13}).
%STEVE: ROMAN TO COMPLETE THIS SECTION INCLUDING THE EFFECT OF STOP MIXING 
%One should also not forget the tree-level condition for no fine-tuning 
%from Eq.\ref{7} that $\mu \lsim v$.

Eq.~(\ref{361}) also indicates that in order to avoid tuning one
has to ensure that the parameter $\mu$ has a reasonably small value.
To avoid tuning entirely one should expect $\mu$ to be less than $M_Z$.
However, so small values of the parameter $\mu$ are ruled out by
chargino searches at LEP. Therefore in our analysis we allow the
effective $\mu_\mathrm{eff}$ parameter to be as large as $200\,\mbox{GeV}$ 
that does not result in enormous fine-tuning.

\section{The NMSSM \label{nmssm}}  
In this paper, we only consider the NMSSM with a scale invariant
superpotential. Alternative models known as the minimal
non-minimal supersymmetric SM (MNSSM), new minimally-extended supersymmetric SM
or nearly-minimal supersymmetric SM (nMSSM) or with additional $U(1)'$ gauge
symmetries have been considered elsewhere \cite{other-non-minimal},
as has the case of explicit CP violation \cite{cpvnmssm}. \sn 

The NMSSM superpotential is given, in terms of (hatted) superfields, by
\beq
{\cal W} = \lambda \widehat{S} \widehat{H}_u \widehat{H}_d +
\frac{\kappa}{3} \, \widehat{S}^3 + h_t
\widehat{Q}_3\widehat{H}_u\widehat{t}_R^c - h_b \widehat{Q}_3
\widehat{H}_d\widehat{b}_R^c  - h_\tau \widehat{L}_3 \widehat{H}_d
\widehat{\tau}_R^c \; ,
\label{supot2}
\eeq
in which only the third generation fermions have been included.
%(with possible neutrino Yukawa couplings have been set to zero), and $\widehat
%Q,  \widehat L$ stand for superfields associated with the $(t,b)$ and
%$(\tau,\nu_\tau)$  SU(2) doublets. 
The first two terms substitute the $\mu
\widehat H_u  \widehat H_d$ term in the MSSM superpotential, while the three
last terms  are the usual generalization of the Yukawa interactions. The soft
SUSY breaking terms consist of  the scalar mass terms for the Higgs and sfermion
scalar fields which, in terms of the  fields corresponding to the complex scalar
components of the superfields, are given by,
\beq
\label{Lmass}
 -{\cal L}_\mathrm{mass} &=& 
 m_{H_u}^2 | H_u |^2 + m_{H_d}^2 | H_d|^2 + m_{S}^2| S |^2 \non \\
  &+& m_{{\tilde Q}_3}^2|{\tilde Q}_3^2| + m_{\tilde t_R}^2 |{\tilde t}_R^2|
 +  m_{\tilde b_R}^2|{\tilde b}_R^2| +m_{{\tilde L}_3}^2|{\tilde L}_3^2| +
 m_{\tilde  \tau_R}^2|{\tilde \tau}_R^2|\; .
\eeq
The trilinear interactions between the sfermion and Higgs fields are, 
\beq
\label{Trimass}
-{\cal L}_\mathrm{tril}=  \lambda A_\lambda H_u H_d S + \frac{1}{3}
\kappa  A_\kappa S^3 + h_t A_t \tilde Q_3 H_u \tilde t_R^c - h_b A_b
\tilde Q_3 H_d \tilde b_R^c - h_\tau A_\tau \tilde L_3 H_d \tilde \tau_R^c
+ \mathrm{h.c.}\;.
\eeq
In the unconstrained NMSSM considered here, with non--universal soft
terms at the GUT scale, the
three SUSY breaking masses squared for $H_u$, $H_d$ and $S$ appearing in ${\cal
L}_\mathrm{mass}$ can be expressed in terms of their VEVs  through the three
minimization conditions of the scalar potential.  Thus, in contrast to the MSSM
(where one has only two free parameters at the tree level, generally chosen to
be the ratio of Higgs vacuum expectation values, $\tan\beta$, and the mass
of the pseudoscalar Higgs boson), the Higgs sector of the NMSSM is described by
the six parameters
\beq
\lambda\ , \ \kappa\ , \ A_{\lambda} \ , \ A_{\kappa} \ , \ 
\tan \beta =\ \langle H_u \rangle / \langle H_d \rangle \ \mathrm{and}
\ \mu_\mathrm{eff} = \lambda \langle S \rangle\; .
\eeq
We follow the sign conventions such that the parameters 
$\lambda$ and $\tan\beta$ are positive, while the parameters $\kappa$, 
$A_\lambda$, $A_{\kappa}$ and $\mu_{\mathrm{eff}}$ can have both
signs. \sn

In addition to these six parameters of the Higgs sector, one needs to specify
the soft SUSY breaking mass terms in Eq.~(\ref{Lmass}) for the scalars, the trilinear
couplings in Eq.~(\ref{Trimass}) as well as the gaugino soft SUSY breaking mass parameters  
to describe the model completely, 
\beq 
-{\cal L}_\mathrm{gauginos}= \frac{1}{2} \bigg[ M_1 \tilde{B}  
\tilde{B}+M_2 \sum_{a=1}^3 \tilde{W}^a \tilde{W}_a +
M_3 \sum_{a=1}^8 \tilde{G}^a \tilde{G}_a  \ + \ {\rm h.c.} 
\bigg].
\eeq
Clearly, in the limit $\lambda \to 0$ with finite $\mu_{\rm eff}$, the NMSSM
turns into the MSSM with a decoupled singlet sector. Whereas the phenomenology
of the NMSSM for  $\lambda \to 0$ could still differ somewhat from the MSSM in
the case  where the lightest SUSY particle  is the singlino   (and hence
with the possibility of a long lived next-to-lightest SUSY particle
\cite{singlsp}),  we will not consider this situation here. 
In fact we shall be interested exclusively in large values of $\lambda$ 
(i.e. $\lambda > 0.5$) in order to increase the tree-level Higgs mass as 
in Eq.~(\ref{eq:hmassNMSSM}). For the same reason we shall also focus on 
moderate values of $\tan\beta$ ($\tan\beta =2,3 $) that result in the relatively
large values of the top quark Yukawa coupling $h_t$ at low energies.  \sn

\vspace*{0.2cm}
\begin{table}[!h]
  \centering
  \begin{tabular}{|c|c|c|c|c|c|c|}
    \hline
$\kappa(M_Z)$& $0$    & $0.1$  & $0.2$ & $0.3$ & $0.4$ & $0.5$\\
 \hline
$\tan\beta=2$& $0.62$ & $0.61$ & $0.60$& $0.58$& $0.53$& $0.42$\\
 \hline
$\tan\beta=3$& $0.68$ & $0.68$ & $0.66$& $0.63$& $0.56$& $0.45$\\
 \hline
%$\tan\beta=4$& $0.7$ & $0.69$ & $0.68$& $0.64$& $0.57$& $0.45$\\
%\hline
  \end{tabular}
  \caption{Two-loop upper bounds on $\lambda(M_Z)$ for different values of $\kappa(M_Z)$
and $\tan\beta$ in the NMSSM.}
  \label{table:lambda1}
\end{table}
\vspace*{0.2cm}
The growth of the Yukawa couplings $h_t$, $\lambda$ and $\kappa$ at the
electroweak (EW) scale 
entails the increase of their values at the Grand Unification scale $M_X$ resulting 
in the appearance of the Landau pole. Large values of $h_t(M_X)$, $\lambda(M_X)$ 
and $\kappa(M_X)$ spoil the applicability of perturbation theory at high 
energies so that the RGEs cannot be used for an adequate description 
of the evolution of gauge and Yukawa couplings at high scales $Q\sim M_X$. 
The requirement of validity of perturbation theory up to the Grand Unification 
scale restricts the interval of variations of Yukawa couplings at the
EW scale. In particular, the assumption that perturbative physics 
continues up to the scale $M_X$ sets an upper limit on the low energy value of 
$\lambda(M_Z)$ for each fixed set of $\kappa(M_Z)$ and $h_t(M_t)$ (or $\tan\beta$). 
With decreasing (increasing) $\kappa(M_Z)$ the maximal possible value of 
$\lambda(M_Z)$, which is consistent with perturbative gauge coupling unification, 
increases (decreases) for each particular value of $\tan\beta$. 
In Table~\ref{table:lambda1} we display two-loop upper bounds on $\lambda(M_Z)$ 
for different values of $\kappa(M_Z)$ and $\tan\beta$ in the NMSSM.
As one can see the allowed range for the Yukawa couplings varies when 
$\tan\beta$ changes. Indeed, for $\tan\beta=2$ the value of $\lambda(M_Z)$
should be smaller than $0.62$ to ensure the validity of perturbation theory up 
to the scale $M_X$. At large $\tan\beta$ the allowed range for the Yukawa 
couplings enlarges. The upper bound on $\lambda(M_Z)$ grows with increasing 
$\tan\beta$ because the top--quark Yukawa coupling decreases. At large 
$\tan\beta$ (i.e. $\tan\beta > 4-5$) the upper bound on $\lambda(M_Z)$ 
approaches the saturation limit where $\lambda_{\max}\simeq 0.72$. \sn

%Clearly, in the limit $\lambda \to 0$ with finite $\mu_{\rm eff}$, the NMSSM
%turns into the MSSM with a decoupled singlet sector. Whereas the phenomenology
%of the NMSSM for  $\lambda \to 0$ could still differ somewhat from the MSSM in
%the case  where the lightest SUSY particle  is the singlino   (and hence
%with the possibility of a long lived next-to-lightest SUSY particle
%\cite{singlsp}),  we will not consider this situation here. 
%In fact we shall be interested exclusively in large values of $\lambda$ in order to 
%increase the tree-level Higgs mass as in Eq.~\ref{eq:hmassNMSSM}. 
%However $\lambda$ cannot be too large otherwise a Landau pole
%is reached below the cut-off scale, which we take to be the unification scale $M_{GUT}$.
%We now discuss how large $\lambda$ can be in the NMSSM.
%In Table~\ref{table:lambda1} we display one-loop upper bounds on
%$\lambda(M_Z)$ for different values of $\kappa(M_Z)$ and $\tan\beta$
%in the NMSSM. The one-loop upper bounds on
%$\lambda(M_Z)$ for different values of $\kappa(M_Z)$ and $\tan\beta$
%in the NMSSM supplemented by three $SO(10)$ $10$--plets, or
%equivalently three $SU(5)$ $5+\overline{5}$--plets are displayed in 
%Table~\ref{table:lambda2} for the case that the masses of all extra
%exotic states are set to be equal to $300\,\mbox{GeV}$ and in
%Table~\ref{table:lambda3} in case they are set equal to $1\,\mbox{TeV}$.
%The two loop renormalisation group equations (RGEs) used to obtain these results 
%are given in Appendix~\ref{sec}.
\vspace*{0.2cm}
\begin{table}[!h]
  \centering
  \begin{tabular}{|c|c|c|c|c|c|c|c|}
    \hline
$\kappa(M_Z)$& $0$    & $0.1$  & $0.2$ & $0.25$ & $0.3$ & $0.4$ & $0.5$\\
 \hline
$\tan\beta=2$& $0.72$ & $0.72$ & $0.7$& $0.67$  & $0.65$& $0.57$& $0.45$\\
 \hline
$\tan\beta=3$& $0.76$ & $0.75$ & $0.73$& $0.70$  & $0.67$& $0.59$& $0.47$\\
 \hline
%$\tan\beta=4$& $0.77$ & $0.76$ & $0.74$& $0.71$ & $0.68$& $0.6$& $0.47$\\
%\hline
  \end{tabular}
  \caption{Two-loop upper bounds on $\lambda(M_Z)$ for different values of $\kappa(M_Z)$
and $\tan\beta$ within the NMSSM supplemented by three $SU(5)$ $(5+\overline{5})$--plets.
The masses of all extra exotic states are set to be equal to $300\,\mbox{GeV}$.}
  \label{table:lambda2}
\end{table}
\vspace*{0.2cm}
\vspace*{0.2cm}
\begin{table}[!h]
  \centering
  \begin{tabular}{|c|c|c|c|c|c|c|c|}
    \hline
$\kappa(M_Z)$& $0$    & $0.1$  & $0.2$ & $0.25$ & $0.3$ & $0.4$ & $0.5$\\
 \hline
$\tan\beta=2$& $0.71$ & $0.7$ & $0.68$& $0.66$  & $0.64$& $0.56$& $0.45$\\
 \hline
$\tan\beta=3$& $0.74$ & $0.74$ & $0.72$& $0.69$  & $0.67$& $0.59$& $0.46$\\
 \hline
%$\tan\beta=4$& $0.76$ & $0.75$ & $0.73$& $0.70$ & $0.68$& $0.60$& $0.47$\\
%\hline
  \end{tabular}
  \caption{Two-loop upper bounds on $\lambda(M_Z)$ for different values of $\kappa(M_Z)$
and $\tan\beta$ within the NMSSM supplemented by three $SU(5)$ $(5+\overline{5})$--plets.
The masses of all extra exotic states are set to be equal to $1\,\mbox{TeV}$.}
  \label{table:lambda3}
\end{table} 
\vspace*{0.2cm}

The renormalisation group (RG) flow of the Yukawa couplings depends rather
strongly on the values of the gauge couplings at the intermediate scales.
To demonstrate this we examine the RG flow of gauge and Yukawa couplings 
within the SUSY model that contains three extra $SU(5)$ $(5+\overline{5})$--plets 
that survive to low energies and can form three $10$--plets in the SUSY-GUT 
model based on the $SO(10)$ gauge group. In this SUSY model the strong gauge 
coupling has a zero one--loop beta function whereas at two--loop level 
the coupling has a mild growth as the renormalisation scale increases.
Since extra states form complete $SU(5)$ multiplets the high-energy scale 
where the unification of the gauge couplings takes place remains almost
the same as in the MSSM. At the same time extra multiplets of matter change 
the running of the gauge couplings so that their values at the intermediate
scale rise substantially. In fact the two--loop beta functions
of the SM gauge couplings are quite close to their saturation limits when 
these couplings blow up at the GUT scale. Further enlargement of the 
particle content can lead to the appearance of the Landau pole during
the evolution of the gauge couplings from $M_Z$ to $M_X$. Because the beta 
functions are so close to the saturation limits the RG flow of the gauge
couplings (i.e. their values at the intermediate scale) also 
depends on the masses of the extra exotic states. Since $g_i(Q)$ occurs in 
the right--hand side of the RGEs for the Yukawa couplings with 
negative sign the growth of the gauge couplings prevents the appearance 
of the Landau pole in the evolution of these couplings. It means that 
for each value of $h_t(M_t)$ (or $\tan\beta$) and $\kappa(M_Z)$
the upper limit on $\lambda(M_Z)$ increases as compared with the NMSSM.
The two-loop upper bounds on $\lambda(M_Z)$ for different values of 
$\kappa(M_Z)$ and $\tan\beta$ in the NMSSM supplemented by three $SO(10)$ 
$10$--plets, or equivalently three $SU(5)$ $(5+\overline{5})$--plets are 
displayed in Table~\ref{table:lambda2} for the case that the masses of
all extra exotic states are set to be equal to $300\,\mbox{GeV}$ and
in Table~\ref{table:lambda3} in case they are set to be equal to
$1\,\mbox{TeV}$. The two-loop RGEs used to obtain these results
are given in Appendix~\ref{sec}. Because the RG flow of the gauge couplings 
depends on the masses of extra  exotic states the upper bounds on 
$\lambda(M_Z)$ presented in Tables \ref{table:lambda2} and \ref{table:lambda3}
are slightly different. The restrictions on $\lambda(M_Z)$ obtained in
this section are useful for the phenomenological analysis which we are going 
to consider next.

%%%%%%%%%%%%%%%%%%%%%%%%%%%%%%%%%%%%%%%%%%%%%%%%%%%%%%%%%%%%%%%%%%%
\section{Constraints from Higgs Boson Searches, SUSY Particle Sear\-ches
  and Dark Matter \label{sec:constraints}}

Our scenarios are subject to constraints on the Higgs boson masses
from the direct searches at LEP, Tevatron and the LHC. Also the SUSY
particle masses have to be compatible with the limits given by the
experiments. Finally, the currently measured value of the relic
density shall be reproduced. Further constraints arise  from the low-energy observables. \sn
 
\subsection{Higgs boson searches}
We start by discussing the constraints which arise from the LHC search
for the Higgs boson. At the LHC, the most relevant Higgs
boson production channels for neutral (N)MSSM Higgs bosons are given
by gluon fusion, gauge boson fusion, Higgs-strahlung and associated
production with a heavy quark pair. The two main mechanisms for
charged Higgs boson production are top quark decay and associated
production with a heavy quark pair. For reviews, see
\cite{Djouadi:2005gj,mssmreviews,higgswg}. As in our scenarios the charged Higgs
boson mass is larger than 450 GeV and hence well beyond the sensitivity
of Tevatron and current LHC searches, we will discuss in the following
only neutral Higgs boson production. \sn

\noindent
\underline{\bf{Gluon fusion}} In the SM and in SUSY extensions, such as
the (N)MSSM, for low values of $\tan\beta$, the most important
production channel is given by gluon fusion \cite{Georgi:1977gs}. In
the NMSSM we have 
\beq 
gg \to H_i \quad \mbox{and} \quad gg \to A_j \;  , \qquad i=1,2,3, \;
j=1,2 \; .
\eeq
Since this is the dominant Higgs production mechanism for a 125 GeV
Higgs boson at the LHC, we find it convenient to define for later use
the ratio of the gluon fusion production cross-section for the Higgs boson
$H_i$ in the NMSSM to the gluon fusion production cross-section for a
SM Higgs boson $H^{SM}$ of same mass as $H_i$, 
\beq  
\label{gg}
R_{\sigma_{gg}}(H_i)\equiv \frac{\sigma(gg\rightarrow
  H_i)}{\sigma(gg\rightarrow H^{SM})} \;.
\eeq
Gluon fusion is mediated by heavy quark loops in the SM and
additionally by heavy squark loops in the (N)MSSM. 
It is subject to important higher-order QCD corrections. For the SM,
they have been calculated at next-to-leading order (NLO) \cite{ggnlo}
including the
full mass dependence of the loop particles and in the heavy top quark
limit, and up to next-to-next-to-leading order (NNLO) in the heavy top
quark limit \cite{ggnnlo}. The cross-section has been improved by
soft-gluon resummation at next-to-next-to-leading logarithmic (NNLL)
accuracy \cite{resum}. Top quark mass effects on the NNLO loop
corrections have been studied in \cite{ozeren}, and the 
EW corrections have been provided in \cite{ewcorr}. In the MSSM, the
QCD corrections have been calculated up to NLO \cite{ggnlo}. The
QCD corrections to squark loops have been first considered in
\cite{squarkfirst} and at full NLO SUSY-QCD in the heavy mass limit in
\cite{nlosqcdheavy}. The
(s)bottom quark contributions at NLO SUSY-QCD have been taken into
account through an 
asymptotic expansion in the SUSY particle masses \cite{asymptotic}. For
squark masses below $\sim 400$ GeV, mass effects play a role and can
alter the cross-section by up to 15\% compared to the heavy mass limit
as has been shown for the QCD corrections to the squark loops in
\cite{qcdsquark,Muhlleitner:2006wx}.  The SUSY QCD corrections including the full mass
dependence of all loop particles have been provided by
\cite{fullsusyqcd}. The mass effects turn out to be
sizeable. The NNLO SUSY-QCD corrections from the (s)top
quark sector to the matching coefficient determining the effective
Higgs gluon vertex have been calculated in \cite{zerf}. \sn

The gluon fusion cross-section has been implemented
in the Fortran code {\tt HIGLU}  \cite{higlu} up to NNLO QCD. While at
NLO the full mass dependence of the loop particles is taken into
account the NNLO corrections are obtained in an effective theory
approach. In the MSSM the full squark mass dependence in the NLO QCD
corrections to the squark loops is included \cite{Muhlleitner:2006wx}. Note, however, that in
the MSSM at NNLO the mismatch in the QCD corrections to the effective
vertex is not taken into account, neither the SUSY QCD corrections to
the effective vertex. The former should be only a minor effect,
though, as the dominant effect of the QCD higher-order corrections stems from
the gluon radiation. Furthermore, the EW corrections to the SM can be
obtained with {\tt HIGLU}. 
In order to check if our scenario is compatible with the recent LHC
results, we need the cross-section of a SM-like Higgs boson of 124 to 126 GeV. The experiments
include in their analyses the NNLO QCD (CMS also the NNLL QCD and NLO
EW) corrections \cite{AtlasTalk,CMSTalk} to the
gluon fusion cross-section as provided by the Higgs Cross-Section 
Working Group \cite{higgswg}. For SUSY, however, the EW corrections
are not available. In order to be consistent, we therefore
compare in the following the NMSSM cross-section to the SM cross-section
at NNLO QCD. As the QCD corrections are not affected by modifications of
the Higgs couplings to the (s)quarks, the NMSSM cross-section can be obtained with
the program {\tt HIGLU} by multiplying the MSSM Higgs couplings with the
corresponding modification factor of the NMSSM Higgs couplings to the
(s)quarks with respect to the MSSM case. We have implemented these
coupling modification factors in the most recent {\tt HIGLU} version 3.11. \sn

\noindent
\underline{\bf{\boldmath{$W/Z$}-boson fusion}} 
Gauge boson fusion \cite{wzfusion} plays an important role for light
CP-even Higgs boson production in the SM limit,\footnote{The
quark $q$ stands for a generic quark flavour, which is different for
the two quarks in case of $W$-boson fusion. The same notation is
applied below in Higgs-strahlung.}
\beq
qq \to qq + W^* W^*/Z^* Z^* \to qq H_i \; , \qquad i=1,2,3 \;.
\eeq
Otherwise the (N)MSSM cross-section is suppressed with respect to the
SM case by mixing angles entering the Higgs couplings to the gauge
bosons. The NLO QCD corrections are of ${\cal O} (10\%)$ of the total cross-section \cite{nloqcdwfusion,fortsch}. The full EW and QCD
corrections to the SM are ${\cal O} (5\%)$
\cite{ciccolini}.  The NNLO QCD effects on the cross-section amount to
$\sim 2$\% 
\cite{wwnnloqcd}. The SUSY QCD and SUSY EW corrections are small
\cite{wwsusyqcdew1,wwsusyqcdew2}. Once again, as QCD corrections
are not affected by the Higgs couplings to the gauge bosons, the QCD
corrected NMSSM gauge boson fusion production cross-sections can be derived from the QCD corrected SM cross-section by simply
applying the modification factor of the respective NMSSM Higgs coupling to the
gauge bosons with respect to the SM coupling,
\beq
\sigma_{QCD}^{NMSSM} (qqH_i) = \left(
  \frac{g_{VVH_i}}{g_{VVH^{SM}}} \right)^2 \sigma_{QCD}^{SM}
(qqH^{SM}) \; , \quad V=W,Z \;,
\eeq
where $g$ denotes the coupling.
The EW corrections, however, cannot be taken over. We have obtained the
SM production cross-section at NLO QCD from the program {\tt VV2H}
\cite{programs}. While the experiments use the SM cross-section at
NNLO QCD (CMS also at NLO EW), the effects of these
additional corrections in the SM limit, where we compare our NMSSM
Higgs cross-section to the SM case, are small. \sn

\noindent 
\underline{\bf Higgs-strahlung} The CP-even Higgs bosons can also be
produced in Higgs-strahlung \cite{higgsrad},
\beq
qq \to VH_i  \; , \qquad  V=W,Z, \quad i=1,2,3 \;,
\eeq
with the NMSSM cross-section always being suppressed by mixing angles
compared to the SM cross-section. The QCD corrections
apply both to the SM and (N)MSSM case. While the NLO QCD corrections
increase the cross-section by ${\cal O} (30\%)$ \cite{fortsch,qcdhrad}
the NNLO QCD corrections are small \cite{nnlohrad}. The full EW
corrections are only known for the SM and decrease the cross-section by
${\cal O} (5-10\%)$ \cite{ewhrad}. The SUSY-QCD corrections amount to
less than a few 
percent \cite{wwsusyqcdew1}.  The NLO QCD SM Higgs-strahlung cross-section
has been obtained with the program {\tt V2HV} \cite{programs}. The
NMSSM Higgs production cross-sections can be derived from it by
applying the Higgs coupling modification factors,
\beq
\sigma_{QCD}^{NMSSM} (VH_i) = \left(
  \frac{g_{VVH_i})}{g_{VVH^{SM}}} \right)^2 \sigma_{QCD}^{SM}
(VH^{SM}) \; , \quad V=W,Z \;.
\eeq  
The experiments use the QCD corrected cross-section up to NNLO
(CMS also including the NLO EW corrections). While we neglect the NNLO and EW
corrections, we do not expect this to influence significantly the total cross-section composed of all production channels, in view of
the small size of the total Higgs-strahlung cross-section itself. \sn

\noindent
\underline{\bf Associated production with heavy quarks} Associated
production of (N)MSSM Higgs bo\-sons with top quarks \cite{lotth} only
plays a role for the light scalar Higgs particle and small values of
$\tan\beta$ due to the suppression of the Higgs couplings to top
quarks $\sim 1/\tan\beta$. While associated production with bottom
quarks \cite{lotth,nlobbh} does not play a role in the SM, in the
(N)MSSM this cross-section becomes important for large values of
$\tan\beta$ and can exceed the gluon fusion cross-section. As our
scenarios include small values of $\tan\beta$ we will not further discuss
this cross-section here. The values for the SM $t\bar{t}H$ cross
section including NLO QCD corrections \cite{nlotth}, which are of
moderate size, can be obtained from the Higgs Cross Section Working
Group homepage \cite{higgswgweb}. From these we derived the NMSSM cross-section values by
replacing the SM Yukawa couplings with the NMSSM Yukawa couplings,
\beq
\sigma_{QCD}^{NMSSM} (t\bar{t} H_i) = \left(
  \frac{g_{t\bar{t} H_i}}{g_{t\bar{t}H^{SM}}} \right)^2 \sigma_{QCD}^{SM}
(t\bar{t}H^{SM}) \;.
\eeq
The NLO SUSY QCD corrections, which have not been taken into account
by the experiments, are of moderate size \cite{sqcdtth} . 

%%%%%%%%%%%%%%%%%%%%%%%%%%%%%%%%%%%%%%%%%%%%%%%%%%%%%%%
\subsection{Constraints from the LHC Searches}
Recent results presented by the ATLAS \cite{AtlasTalk} and the CMS
\cite{CMSTalk} Collaborations seem to indicate a Higgs boson of
mass of 126 and 124 GeV, respectively. Based on the dataset corresponding
to an integrated luminosity of up to 4.9 fb$^{-1}$ collected at
$\sqrt{s}=7$ TeV, an excess of events is observed by the ATLAS
experiment for a Higgs boson mass hypothesis close to $M_H=126$ GeV
with a maximum local significance of 3.6$\sigma$ above the expected SM
background. The three most sensitive channels in this mass range are
given by $H\to \gamma \gamma$, $H \to ZZ^{(*)}\to l^+ l^- l^+ l^-$ and
$H\to WW^{(*)} \to l^+\nu l^-\bar{\nu}$.  The CMS Collaboration
presented results of SM Higgs boson searches in the mass range 100-600
GeV in 5 decay modes, $H\to\gamma \gamma, bb, \tau\tau, WW$ and
$ZZ$. The data correspond to an integrated luminosity of up to 4.7
fb$^{-1}$ at $\sqrt{s}=7$ TeV. A modest excess of events is observed
for Higgs boson mass hypotheses towards the low end of the
investigated Higgs 
mass range.  The maximum local significance amounts to 2.6$\sigma$ for
a Higgs boson mass hypothesis of $M_H=124$ GeV. For our NMSSM
benchmark scenarios presented below to be consistent with these LHC
results we demand the production cross-section of the SM-like NMSSM
Higgs boson with mass 124 GeV to 126 GeV (depending on the scenario)
to be compatible within 20\% with the production cross-section of a SM
Higgs boson of same mass. The 20\% are driven by the theoretical
uncertainty on the inclusive Higgs production cross-section given by
the sum of the most relevant production channels at low values of
$\tan\beta$, i.e. gluon fusion, weak boson fusion,
Higgs-strahlung and $t\bar{t}$ Higgs production. The theoretical error
is largest for the gluon fusion cross-section with 10-15\% at these Higgs mass
values and $\sqrt{s}=7$ TeV \cite{higgswg}, and which contributes
dominantly to the inclusive production. We do not consider any
experimental error since this is beyond our scope.  \sn

For simplicity, and
since these search channels are common to both experimental analyses,
we consider the Higgs decays into $\gamma\gamma$, $ZZ \to 4l$ and $WW\to 2l
2\nu$. In order to get an estimate of how closely the NMSSM Higgs resembles the 
SM Higgs in LHC searches in these channels we define the ratios of
branching ratios into massive gauge boson final states
$VV$, where $V=W,Z$,\footnote{The ratio is the same for $WW$ and $ZZ$ final states,
  respectively, as the NMSSM coupling to $WW$ and $ZZ$ is suppressed by the
  same factor compared to the SM.}
and into $\gamma\gamma$, respectively, for an NMSSM Higgs
boson $H_i$ and the SM Higgs boson $H^{SM}$ of same mass,
\beq 
\label{VV}
R_{VV}(H_i)\equiv \frac{BR(H_i\rightarrow VV)}
{BR(H^{SM}\rightarrow VV)} \quad \mbox{ and } \quad
R_{\gamma\gamma}(H_i)\equiv \frac{BR(H_i\rightarrow \gamma\gamma)}
{BR(H^{SM}\rightarrow \gamma\gamma)} \;. 
\eeq
We also define analogously $R_{\Gamma_{tot}}$ for the total widths,
\beq 
\label{rattot}
R_{\Gamma_{tot}} (H_i)\equiv \frac{\Gamma_{tot}(H_i)}
{\Gamma_{tot}(H^{SM})} \;  ,
\eeq
and $R_{b\overline{b}}(H_i)$ for the decay into $b\overline{b}$. 
Although this final state is not useful for LHC searches, it is interesting to show as in
this mass range and for small values of $\tan\beta$ the decay into
$b\bar{b}$ contributes dominantly to the total width. Depending on of how
much $H_d$ component is in the mass eigenstate $H_i$ it is enhanced or
suppressed compared to the SM. This directly influences the total
width and hence the branching ratios of the other final states. \sn

For a crude estimate of the total Higgs cross-section at the LHC,
we can combine these channels in quadrature in contrast to the
experiments which do a sophisticated statistical combination of the
various search channels, which is, however, beyond the scope of our
theoretical analysis. Our results should therefore only be regarded as 
a rough estimate which is indicative enough, however, at the present
status of the experimental research, to exclude or not exclude a benchmark
scenario. We hence demand for a scenario to be valid that one of the
NMSSM Higgs bosons $H_i$ satisfies
\beq
0.8\, \sigma_{tot}(H^{SM}) \le \sigma_{tot}(H_i) \le 1.2\,
\sigma_{tot} (H^{SM})\;,
\label{eq:cxnconstraint}
\eeq
where
\beq
\sigma_{tot}(H)&=& \sigma_{incl}(H)\Big\{ BR^2 (H \to \gamma \gamma)
+ 16\, BR^2 (H \to ZZ) \, BR^2 (Z \to ll) \, BR^2 (Z \to ll) \nonumber \\ 
&+& 16\, BR^2 (H \to WW) \, BR^2 (W\to l\nu) \, BR^2 (W\to l\nu) \Big\}^{1/2}
\label{eq:sigtot} \;.
\eeq
The inclusive cross-section $\sigma_{incl}$ is composed of gluon
fusion, vector boson fusion, Higgs-strahlung and associated production
with $t\bar{t}$,
\beq
\sigma_{incl}(H) = \sigma (gg\to H) + \sigma (Hqq) + \sigma (WH) +
\sigma (ZH) +\sigma (t\bar{t}H) \; , 
\eeq
with $H=H_i,H^{SM}$, respectively, subject to the constraint
$M_{H^{SM}}=M_{H_i}= 124$-126 GeV, depending on the scenario under
consideration. It is dominated by the gluon fusion cross-section.
The factors 16 in Eq.~(\ref{eq:sigtot}) arise from the
sum of the four possible lepton final states in the decays of the $Z$-
and $W$-boson pairs, respectively. (We neglect interference effects.) 
For the gauge boson branching
ratios we chose the values given by the Particle Data Group
\cite{pdg},
\beq
BR(Z\to ll) = 0.0366 \; , \qquad BR(W\to l\nu) = 0.1080 \;.
\eeq
It is useful to define
\be
\label{tot}
R_{\sigma_{tot}}(H_i) =\frac{\sigma_{tot} (H_i)}{\sigma_{tot} (H^{SM})},
\ee
in order to provide a measure of how closely the NMSSM Higgs resembles the 
SM Higgs in the most important current LHC search channels.
The total cross-section is dominated by the $WW$-boson final
state due to the large branching ratio. As we will see below, in the
NMSSM the branching ratio into $\gamma\gamma$ can be enhanced for
certain parameter configurations compared to the SM. To illustrate
this effect, we therefore also calculate separately the ratios of the expected
number of events in the NMSSM compared to the SM for the
$\gamma\gamma$ final state and for the $VV$ ($V=W,Z$)
final state, which is the same for $V=W$ or $Z$. They are given by 
\beq
R_{\sigma_{incl}}(H_i)  R_{\gamma\gamma}(H_i) \quad \mbox{and} \quad  
R_{\sigma_{incl}}(H_i)  R_{VV}(H_i) 
\; ,
\label{eq:separate}
\eeq
where $R_{\sigma_{incl}}(H_i)=\sigma_{incl}(H_i)/\sigma_{incl}
(H^{SM})$. 

%%%%%%%%%%%%%%%%%%%%%%%%%%%%%%%%%%%%%%%%%%%%%%%%%%%%%%

\subsection{NMSSM spectrum and NMSSM Higgs boson branching ratios}
The SUSY particle and NMSSM Higgs boson masses and branching ratios
are calculated with the program package {\tt NMSSMTools}
\cite{nmhdecay1,nmhdecay2,nmssmtools}. As for the NMSSM Higgs boson
masses, the leading one-loop contributions due to heavy (s)quark loops
calculated in the effective potential approach \cite{effpotnmssm}, the
one-loop contributions due to chargino, neutralino and scalar loops in leading
logarithmic order in Ref. \cite{llognmssm} and the leading logarithmic
two-loop terms of ${\cal O} (\alpha_t \alpha_s)$ and ${\cal O}
(\alpha_t^2)$, taken over from the MSSM results, have been implemented
in {\tt NMSSMTools}. The full one-loop contributions have
been computed in the $\overline{\mbox{DR}}$ renormalisation scheme
\cite{slavich,spectrum3a} and also in a mixed on-shell (OS) and
$\overline{\mbox{DR}}$ scheme as well as in a pure OS scheme
\cite{ender}. Furthermore, the ${\cal O} (\alpha_t \alpha_s + \alpha_b 
\alpha_s)$ corrections have been provided in the approximation of zero
external momentum \cite{slavich}. The corrections provided by
Ref.~\cite{slavich} have been implemented in {\tt NMSSMTools} as well. \sn

The calculation of the NMSSM Higgs boson decay widths and
branching ratios within {\tt NMSSMTools} is performed by the Fortran
code {\tt NMHDECAY} \cite{nmhdecay1,nmhdecay2} which uses to some
extent parts of the Fortran code {\tt HDECAY} \cite{hdecay,susyhit} that
calculates SM and MSSM Higgs boson partial widths and branching
ratios. The calculation of the SUSY particle branching ratios on the other hand 
with the Fortran code {\tt NMSDECAY} \cite{nmsdecay} is based on a
generalisation of the Fortran code {\tt SDECAY} \cite{sdecay,susyhit}
to the NMSSM case. {\tt NMSSMTools} provides the output for the
complete NMSSM particle spectrum and mixing angles and for the decays
in the SUSY Les Houches format \cite{slha}. The latter can be read in
by our own Fortran version for NMSSM Higgs boson decays based on an
extension of the latest {\tt HDECAY} version. It reads in the particle
spectrum and mixing angles created with {\tt NMSSMTools}, calculates
internally the NMSSM Higgs boson couplings and uses them to calculate the
Higgs decay widths and branching ratios. The results for the branching
ratios from {\tt NMSSMTools} and from our own program agree reasonably well
and the differences in the total cross-section Eq.~(\ref{eq:sigtot}),
obtained with the results from the two programs, due
to deviations in the branching ratios are in the percent range.

%%%%%%%%%%%%%%%%%%%%%%%%%%%%%%%%%%%%%%%%%%%%%%%%%%%%%%%
\subsection{Constraints from Dark Matter, Low Energy Observables, LEP and Tevatron}

Based on an interface between {\tt NMHDECAY} and {\tt micrOMEGAs}
\cite{micromegas} the relic abundance of the NMSSM dark matter
candidate $\tilde{\chi}_1^0$ can be evaluated using {\tt NMSSMTools}.
As an independent check, we also used the stand alone code {\tt
  micrOMEGAs} to calculate the relic density. All the relevant cross-sections for the lightest neutralino annihilation and co-annihilation are
computed. The density evolution equation is numerically solved and the relic
density of $\tilde{\chi}^0_1$ is calculated. The differences in the
result for the relic density calculated with both tools are negligible. 
The results are compared with
the ``WMAP" constraint  $0.094 \, \lesssim \Omega_{\rm CDM} h^2
\lesssim 0.136$ at the $2\sigma$ level \cite{wmap}.  \sn

When the spectrum and the couplings of the Higgs and SUSY particles are 
computed with {\tt NMSSMTools}, constraints from low-energy
observables  as well as available Tevatron and LEP constraints are checked. The
results of the four LEP collaborations, combined by the LEP Higgs
Working Group, are included \cite{lep}. More specifically, the
following experimental constraints are taken into account: 

\begin{itemize}
\vspace*{-2mm}

\item[(i)]  The masses of the neutralinos as well as their couplings to the $Z^0$
boson are compared with the LEP constraints from direct searches and from the
invisible $Z^0$ boson width. \vspace*{-2mm}

\item[(ii)] Direct bounds from LEP and Tevatron on the masses of the charged
particles ($h^\pm$, $ \chi^\pm$, $\tilde q$,~$\tilde l$) and the gluino are
taken into account. \vspace*{-2mm}

\item[(iii)] Constraints on the Higgs production rates from all channels studied
at LEP. These include in particular $Z H_i$ production, $H_i$ being any of
the  CP--even Higgs particles,  with all possible two body decay modes of
$H_i$ (into $b$ quarks, $\tau$ leptons, jets, photons or invisible), and all
possible decay modes of $H_i$ of the form $H_i \to A_j A_j$,  $A_j$
being any of the  CP--odd Higgs particles, with all possible combinations of
$A_j$ decays into $b$ quarks, $c$ quarks, $\tau$ leptons and jets. Also
considered is the associated production mode $e^+e^- \to H_i A_j$ with,
possibly, $H_i \to A_j A_j$. (In practice, for our purposes, only
combinations of $i=1,2$ with $j=1$ are phenomenologically relevant.)
%{\bf Maggie: Does the statement in brackets only concern the latter
%  decay? Otherwise I am not sure about it.} STEVE: IT ONLY CONCERNS THE LATTER

\item[(iv)] Experimental constraints from B physics \cite{bphys} such as  the
branching ratios of the rare decays BR$(B \to X_s \gamma)$,  BR$(B_s
\to \mu^+ \mu^-)$ and BR$(B^+ \to \tau^+ \nu_\tau)$ and the mass
differences $\Delta M_s$ and $\Delta M_d$, are also implemented; 
compatibility of each point in parameter space with the current
experimental bounds is required at the two sigma level.
\vspace*{-2mm}
\end{itemize}

%%%%%%%%%%%%%%%%%%%%%%%%%%%%%%%%%%%%%%%%%%%%%%%%%%%%%%%
\subsection{Constraints on SUSY particle masses from the LHC}
The ATLAS \cite{atlsearch} and CMS \cite{cmssearch} searches in final
states with jets and missing transverse energy $E_T^{miss}$, with large jet multiplicities and
$E_T^{miss}$, with heavy flavour jets and $E_T^{miss}$ within
simplified models and mSUGRA/constrained MSSM (CMSSM) models set
limits on the masses of gluino and squark masses. Further constraints
are obtained from searches in final states with leptons and taus \cite{wlepton}. The
precise exclusion limits depend on the investigated final state, the value of the neutralino
and/or chargino masses and the considered model. Light gluino (below
about 600 GeV) and squark masses (below about 700 GeV) are
excluded. The limits cannot be applied, however, to the third
generation squarks in a model-independent way. Recent 
analyses scanning over the physical stop and sbottom masses and
translating the limits to the third generation squark sector have shown
that the sbottom and stop masses can still be as light as $\sim 200-300$ GeV
depending on the details of the spectrum \cite{desaietal}. Especially
scenarios, where the lightest stop is the next-to-lightest SUSY
particle (NLSP) and nearly degenerate with the lightest neutralino
assumed to be the lightest SUSY particle (LSP), are challenging for
the experiments. Such scenarios can be consistent with Dark Matter
constraints due to possible co-annihilation \cite{coannihi}. If the
$\tilde{t}_1-\tilde{\chi}_1^0$ mass difference is small enough, the
flavour-changing neutral current decay $\tilde{t}_1 \to c
\tilde{\chi}_1^0$\cite{fcncdec} is dominating and can compete with the four-body
decay into the LSP, a $b$ quark and a fermion pair
\cite{fourbody}. Limits have been placed by the Tevatron searches
\cite{tevstop} and depending on the neutralino mass still allow for
very light stops down to 100 GeV. The authors of Ref.~\cite{stoplhclim}
found that translating the LHC limits to stop searches in the co-annihilation 
scenario \cite{stopsearchcoannihi}\footnote{For stop searches in
  scenarios with light gravitinos see \cite{stopgmsb}.} allows for
stops lighter than $\sim 400$ GeV down to 160 GeV. We are not aware,
however, of any dedicated LHC analysis which excludes very light stops.

%%%%%%%%%%%%%%%%%%%%%%%%%%%%%%%%%%%%%%%%%%%%%%%%%%%%%%%%%%%%%%%%%%%
\section{The benchmark points \label{bm}} 

In this section we present benchmark points for the NMSSM with a
SM-like Higgs boson near 125 GeV. The Higgs sector of the NMSSM has a
rich parameter space including  $\lambda,
\kappa,A_\lambda,A_\kappa,\tan\beta$ and the effective $\mu$ 
parameter. According to the SLHA format \cite{slha}, these parameters
are understood as running $\overline{\mbox{DR}}$ parameters taken at
the SUSY scale $\tilde{M}=1$ TeV while $\tan\beta$ is taken at the
scale of the $Z$ boson mass, $M_Z$. In order not to violate tree-level
naturalness we set $\mu_{eff} \le 200$~GeV for all the considered  
points. The Higgs sector is strongly influenced by the stop sector via radiative
corrections where we further
need to specify the soft SUSY breaking masses $M_{\tilde{Q}_{3}},M_{\tilde{t}_R}$ and the
mixing parameter $X_t$ defined in Eq.~(\ref{Xt}). The main advantage
of the NMSSM over the MSSM, regarding a SM-like Higgs boson near 125 GeV, is
that the stop masses are allowed to be much lighter, making the NMSSM
much more technically natural than the MSSM. Thus all of the
benchmarks discussed here will involve relatively light stops, with
masses well below 1 TeV. We choose low values of $\tan \beta =2,3$ in
order to maximise the tree-level contribution to the Higgs boson mass,
allowing the stops to be lighter.  For very light stops, in order not to be in
conflict with the present exclusion limits, the difference between the
$\tilde{t}_1$ and $\tilde{\chi}_1^0$ masses should be less than $\sim
20-30$ GeV to be in the co-annihilation region. By choosing the
right-handed stop to be the lightest top squark the sbottoms are still
heavy enough to fulfill the LHC limits of about 300 GeV \cite{sbotlim} also in this case. 
On the other hand, the remaining squarks and sleptons 
may be heavier without affecting fine-tuning. In order to satisfy in particular the
LHC search limits for the squarks of the first two generations, 
we set all their masses to be close to 1 TeV and, for simplicity, also
those of the sleptons. 
%Note, however, that our Higgs mass results are not 
%influenced by the exact mass values of these particles. 
To be precise, for the first and second squark and slepton families and for the stau
sector we always take the soft SUSY breaking masses and trilinear
couplings to be 1 TeV, and furthermore, the right-handed soft SUSY
breaking mass and trilinear coupling of the sbottom sector is set to 1
TeV, i.e. ($U\equiv u,c$, $D\equiv d,s$, $E\equiv e,\mu,\tau$)
\beq
M_{\tilde{U}_R}&\!\!=\!\!&
M_{\tilde{D}_R}=M_{\tilde{b}_R}=M_{\tilde{Q}_{1,2}}= 
M_{\tilde{E}_R}  = M_{\tilde{L}_{1,2,3}}=
1 \; \mbox{TeV},
\nonumber \\ 
 A_{U} &\!\!=\!\!& A_{D} = A_b = A_E = 1 \; \mbox{TeV}  \;.
\eeq
This results in physical masses of $\sim$ 1 TeV for the first and
second family squarks and sleptons as well as the heavier sbottom.
These masses can readily be increased without affecting the properties of the
quoted benchmark points appreciably.
The gaugino mass parameters have been set such that they fulfill
roughly GUT relations. Special attention has been paid, however, not
to choose the gluino mass too heavy in order to avoid fine-tuning.
Before discussing the benchmark points, a few technical preliminaries are in order.
The masses for the Higgs bosons and SUSY particles have been obtained
using {\tt NMSSMTools}.  In the presentation of our benchmark scenarios, for the
SM-like Higgs boson we furthermore include the 
ratios of the branching ratios, {\it cf.} Eq.~(\ref{VV}), into
$\gamma\gamma$, $b\bar{b}$ and $VV$ ($V=Z,W$)
as well as the ratio of  the total widths Eq.~(\ref{rattot})
for the NMSSM and SM Higgs boson having the same mass. The NMSSM
branching ratios and total width 
have been obtained with {\tt NMSSMTools} and cross-checked against a private
code whereas the SM values have been calculated with {\tt
  HDECAY}. The latter are shown separately in Table \ref{table:smbr} for a SM Higgs boson with
the mass values corresponding to the various masses of the SM-like
NMSSM Higgs boson in our benchmark scenarios. The NMSSM values can be
obtained by multiplication with the corresponding ratios presented in the
benchmark tables, although it is mainly their relative values,
compared to the SM, that concern us here. Note that {\tt HDECAY}
includes the double off-shell decays 
into massive vector bosons, whereas {\tt NMSSMTools} does not. We
therefore turned off the double off-shell decays also in {\tt HDECAY}. This explains
why the SM values given in Table \ref{table:smbr} differ from the
values given on the website of the Higgs Cross Section Working Group
\cite{higgswgweb}. There are further differences between the two programs in the calculation of
the various partial widths. Thus {\tt HDECAY} includes the full NNNLO
corrections to the top loops in the decay into gluons. Also it uses
slightly different running bottom and charm quark masses. Taking all
these effects into account we estimate the theoretical error on the
ratios of branching ratios, which are calculated with these two
different programs, to be of the order of 5\%. This should be kept in
mind when discussing the benchmark scenarios.
%%%%%%%%%%%%%%%%%%%%%%%%%%%%%%%%%%%%%%%%%%%%%%%%%%%%%%%%%%
\renewcommand{\arraystretch}{1.15}
\begin{table}[!h]
\caption{Branching ratios into $\gamma\gamma$, $ZZ$, $WW$, $b\bar{b}$
  and total width of a SM Higgs boson of mass between 123.5 and 126.5 GeV.}
\vspace*{-6mm}
\label{table:smbr}
\vspace{3mm}
\footnotesize
\begin{center}
\begin{tabular}{|l|r|r|r|r|r|}
\hline
$M_H$ [GeV] & BR($H\to \gamma\gamma$) & BR($H\to ZZ$) & BR($H\to WW$) &
BR($H\to b\bar{b}$) & $\Gamma_{tot}$ [GeV]
\\\hline\hline
123.5 & $2.334 \cdot 10^{-3}$ & 0.018 & 0.174 & 0.616 & $3.773 \cdot 10^{-3}$
\\\hline  
123.6 & $2.335 \cdot 10^{-3}$ & 0.018 & 0.175 & 0.615 & $3.784 \cdot 10^{-3}$
\\\hline  
123.7 & $2.336 \cdot 10^{-3}$ & 0.019 & 0.177 & 0.613 & $3.795 \cdot 10^{-3}$
\\\hline  
123.8 & $2.337 \cdot 10^{-3}$ & 0.019 & 0.179 & 0.612 & $3.807 \cdot 10^{-3}$
\\\hline 
%123.9 & $2.269 \cdot 10^{-3}$ & 0.024 & 0.199 & 0.592 & $3.93 \cdot 10^{-3}$
%\\\hline 
124 & $2.338 \cdot 10^{-3}$ & 0.019 & 0.182 & 0.609 & $3.830 \cdot 10^{-3}$
\\\hline 
124.5 & $2.342 \cdot 10^{-3}$ & 0.020 & 0.189 & 0.601 & $3.890 \cdot 10^{-3}$
\\\hline 
124.6 & $2.343 \cdot 10^{-3}$ & 0.021 & 0.191 & 0.600 & $3.903 \cdot 10^{-3}$
\\\hline 
125 & $2.345 \cdot 10^{-3}$ & 0.021 & 0.197 & 0.594 & $3.953 \cdot 10^{-3}$
\\\hline 
125.8 & $2.347 \cdot 10^{-3}$ & 0.023 & 0.210 & 0.581 & $4.058 \cdot 10^{-3}$
\\\hline 
%125.9 & $2.273 \cdot 10^{-3}$ & 0.029 & 0.231 & 0.561 & $4.206 \cdot 10^{-3}$
%\\\hline
126 & $2.348 \cdot 10^{-3}$ & 0.024 & 0.213 & 0.578 & $4.085 \cdot
10^{-3}$\\ \hline
126.2 & $2.348 \cdot 10^{-3}$ & 0.024 & 0.217 & 0.575 & $4.113 \cdot
10^{-3}$\\ \hline
%126.4 & $2.271 \cdot 10^{-3}$ & 0.030 & 0.240 & 0.553 & $4.281 \cdot 10^{-3}$\\ \hline
126.5 & $2.348 \cdot 10^{-3}$ & 0.025 & 0.222 & 0.570 & $4.155 \cdot 10^{-3}$\\ \hline
\end{tabular}\end{center}
\end{table}
%%%%%%%%%%%%%%%%%%%%%%%%%%%%%%%%%%%%%%%%%%%%%%%%%%%%%%%%%%%%%%%%%

As mentioned above, in order to be compatible with the recent LHC results
for the Higgs boson search we demand the total cross-section as
defined in Eq.~(\ref{eq:sigtot}) to be within 20~\% equal to the
corresponding SM cross-section. We therefore include in our tables for
the benchmark points the ratio $R_{\sigma_{tot}}$, Eq.~(\ref{tot}), of the
NMSSM and SM total cross-section for $\sqrt{s}=7$ TeV and for
completeness also the ratio $R_{\sigma_{gg}}$, Eq.~(\ref{gg}), of the NMSSM and 
SM gluon fusion cross-sections since gluon fusion is the dominant
contribution to inclusive Higgs production for low values of
$\tan\beta$ at the LHC. The latter has been obtained with the Fortran code {\tt
 HIGLU} at NNLO QCD and includes the squark loop
contributions. Note, that we did not include electroweak
corrections. Furthermore, we explicitly verified that $R_{\sigma_{tot}}$ is 
practically the same using NLO or NNLO QCD gluon fusion cross-sections.
The values of the gluon fusion cross-section and of the total cross-section at NLO and NNLO QCD are shown separately for the SM in Table
\ref{table:smcxn}. However, the ratios for the cross-sections which we present in the tables, $R_{\sigma_{gg}}$,
$R_{\sigma_{tot}}$, are for the case of gluon fusion production at
NNLO QCD. With the total cross-section being dominated by the $WW$-boson final
state,  in order to illustrate interesting effects in the NMSSM branching
ratios compared to the SM ones, for the SM-like Higgs boson 
we also give separately the ratios of the
expected number of events in the $\gamma\gamma$ and $VV$ ($V=W,Z$) final states
as given by Eq.~(\ref{eq:separate}). \sn

All our presented  scenarios fulfill the cross-section constraint
Eq.~(\ref{eq:cxnconstraint}) and are hence compatible with the
LHC searches, keeping in mind though that as theorists we can do only a rough
estimate here. Furthermore, they fulfill the constraints from low-energy parameters
as specified in section \ref{sec:constraints} and are compatible with
the measurement of the relic density. In principle they could also account for 
the $3\sigma$ deviation of the muon anomalous magnetic momentum from
the SM if we were to lower the smuon mass, but for clarity we have 
taken all first and second family squark and slepton masses to be
close to 1 TeV, as discussed above. \sn

We consider four different sets of NMSSM benchmark points as follows:
\begin{itemize}
\item
\underline{\bf NMSSM with Lightest Higgs being SM-like near 125 GeV} 

This is achieved with $\lambda = 0.57 - 0.64$ and $\kappa =
0.18-0.25$ such that $\lambda$  does not blow up below the GUT scale
in the usual NMSSM with no extra matter as in Table~\ref{table:lambda1}. 
This set of benchmarks is displayed in Table~\ref{table:nmp1nmp2nmp3}.

\item
\underline{\bf NMSSM with Second Lightest Higgs being SM-like near 125 GeV} 

This is achieved with $\lambda = 0.55 - 0.67$ and $\kappa = 0.10-0.31$
such that $\lambda$ 
does not blow up below the GUT scale in the usual NMSSM with no extra
matter as in Table~\ref{table:lambda1}. 
This set of benchmarks is displayed in Table~\ref{table:nmp4nmp5nmp6}.

\item
\underline{\bf NMSSM-with-extra-matter and Second Higgs being SM-like near 125 GeV} 

This is achieved with $\lambda = 0.68 - 0.69$ and $\kappa = 0.06-0.20$ such that $\lambda$ 
does not blow up below the GUT scale in the NMSSM supplemented by 
three $SO(10)$ $10$--plets as in Table~\ref{table:lambda3}. 
The slightly larger value of $\lambda$ here allows extra matter to be
at or above the TeV scale. 
This set of benchmarks is displayed in Table~\ref{table:nmp7nmp8nmp9}.

\item
\underline{\bf NMSSM-with-extra-matter and Lightest Higgs being SM-like near 125 GeV} 

This is achieved with $\lambda = 0.7$ and $\kappa = 0.20-0.25$ such that $\lambda$ 
does not blow up below the GUT scale in the NMSSM supplemented by
three $SO(10)$ $10$--plets as in Table~\ref{table:lambda2}. The larger
values of $\lambda$ and $\kappa$ here requires the extra matter to be
close to the electroweak scale. This set of benchmarks is displayed in
Table~\ref{table:nmp10nmp11nmp12}. 
\end{itemize}

%%%%%%%%%%%%%%%%%%%%%%%%%%%%%%%%%%%%%%%%%%%%%%%%%%%%%%%%%%%
\renewcommand{\arraystretch}{1.15}
\begin{table}[!h]
\caption{The NLO and NNLO cross-sections for gluon fusion into a SM
  Higgs boson and the total cross-section as defined in
  Eq.~(\ref{eq:sigtot}) including an NLO and NNLO gluon fusion cross-section, respectively, for various values of the SM
  Higgs boson mass.}
\vspace*{-6mm}
\label{table:smcxn}
\vspace{3mm}
\footnotesize
\begin{center}
\begin{tabular}{|l|r|r|r|r|}
\hline
$M_H$ [GeV] & $\sigma^{NLO}_{gg}$ [pb] & $\sigma^{NNLO}_{gg}$ [pb] & $\sigma^{NLO}_{tot}$ [pb] &
$\sigma^{NNLO}_{tot}$ [pb] 
\\\hline\hline
123.5 & 13.08 & 15.44 & 0.134 & 0.155
\\\hline 
123.6 & 13.06 & 15.41 & 0.135 & 0.156
\\\hline 
123.7 & 13.04 & 15.39 & 0.136 & 0.157
\\\hline 
123.8 & 13.02 & 15.36 & 0.137 & 0.158
\\\hline 
124 & 12.97 & 15.31 & 0.138 & 0.160
\\\hline 
124.5 & 12.87 & 15.18 & 0.143 & 0.165
\\\hline 
124.6 & 12.85 & 15.15 & 0.144 & 0.166  
\\\hline 
125 & 12.76 & 15.05 & 0.147 & 0.170
\\\hline 
125.8 & 12.59 & 14.85 & 0.154 & 0.178
\\\hline 
126 & 12.55 & 14.80 & 0.156 & 0.180 \\ \hline
126.2 & 12.51 & 14.75 & 0.158 & 0.182 \\ \hline
126.5 & 12.45 & 14.68 & 0.161 & 0.185 \\ \hline
\end{tabular}\end{center}
\end{table}
%%%%%%%%%%%%%%%%%%%%%%%%%%%%%%%%%%%%%%%%%%%%%%%%%%%%%%%%%%%%%%%%%
Note that the input values for $\lambda$ and $\kappa$ are taken at the
scale 1 TeV. Their corresponding values at $M_Z$ are lower so that
we are well within the limits given in Tables~\ref{table:lambda1}-\ref{table:lambda3}.
Within each set we have selected three benchmark points which are
chosen to illustrate key features of the NMSSM Higgs boson which could be
used to resolve it from the SM Higgs state in future LHC searches. 
This makes 12 benchmark points in total which we label as NMP1 to NMP12.
In the following we discuss the key features of each of the points in
detail under the above four headings. 
Since we focus on large values of $\lambda \gsim 0.55$ and moderate 
values of $\tan\beta =2,3$ the requirement of the validity of perturbation theory up 
to the GUT scale constrains the low energy value of the coupling to be $\kappa\lsim 0.3$. As a result 
the heaviest CP--even, CP--odd and charged Higgs states are almost
degenerate and substantially heavier than the two lightest Higgs scalars 
and the lightest Higgs pseudoscalar for all the benchmark points,
as discussed in \cite{Nevzorov:2004ge}.
Also note that all the benchmark points satisfy the WMAP relic
abundance. Depending on the decomposition of the lightest neutralino
and its mass value the main annihilation channels are
$\tilde{\chi}_1^0\tilde{\chi}_1^0\rightarrow W^+W^-$ or
$\tilde{\chi}_1^0\tilde{\chi}_1^0\rightarrow q\bar{q}$
($q=b,s,d,c,u$). For very light stops $\tilde{t}_1$ with mass near the
$\tilde{\chi}_1^0$ mass, as in the benchmark point NMP8, the dominant
channel is the coannihilation channel,
$\tilde{t}_1 \tilde{\chi}_1^0\rightarrow W^+ b$.

\subsection{\underline{NMSSM with Lightest Higgs being SM-like}}
Table~\ref{table:nmp1nmp2nmp3} shows three points NMP1, NMP2, NMP3
which satisfy NMSSM perturbativity up to the GUT scale
and lead to the SM-like Higgs boson always being the lightest one
$H_1$ near 125~GeV. The mass of the second lightest Higgs boson $H_2$
ranges between 129 and 155 GeV. Being mostly singlet-like its
couplings to SM particles are suppressed enough not to represent any
danger with respect to the LHC Higgs exclusion limits in this mass range.
The benchmark points are ordered in terms of increasing stop mixing $X_t$ scaled
by the geometric mean $m_{\tilde{t}}$ of the soft SUSY breaking stop
mass parameters. The gluino mass is about 700 GeV for all the points and recall that the
squark and slepton masses not shown are all about 1 TeV. \sn

\noindent
\underline{\bf NMP1 with $(\lambda , \kappa) = (0.64,0.25)$} and  $\tan \beta =3$ is
an example of a benchmark point where a SM-like Higgs boson with mass
124.5 GeV can be achieved in the pure NMSSM with relatively small
mixing, $X_t/m_{\tilde{t}}=1.74$. Due to a slightly larger coupling of the
SM-like Higgs boson to down-type quarks in this scenario, the decay into $b\bar{b}$ is
enhanced compared to the SM and also the corresponding branching ratio
(normalised to the SM) of $R_{b\bar{b}}=1.08$. As the decay into $b\bar{b}$
contributes dominantly to the total width, its enhanced value leads to smaller branching ratios
into $ZZ,WW$, $R_{VV}=0.94$ ($V=W,Z$). The partial width into
photons, however, is enhanced compared to the SM due to the additional
SUSY particle loops contributing to the
Higgs-$\gamma\gamma$ coupling with the main enhancement induced by the
chargino loops, so that in the end the branching ratio into photons is
slightly larger than in the SM, $R_{\gamma \gamma }=1.03$. 
As for the gluon fusion cross-section, which is dominated
by the (s)top quark loops at small $\tan\beta$, it is almost the same as in the SM, with
$R_{\sigma_{gg}}=0.97$. This is because the $H_1$ couples with SM-strength
to the top quarks and the effect of the additional squark loops is small. Due to the smaller
branching ratios into $VV$, in particular the dominating one into
$WW$, the total Higgs production cross-section is estimated to be
smaller than in the SM, $R_{\sigma_{tot}}=0.92$, which is compatible
with LHC constraints. The heavier stop mass here of 782 GeV is a little uncomfortable from 
the point of view of fine-tuning, but not too bad. \sn
\begin{table}[!h]
\caption{NMSSM benchmark points with the lightest Higgs boson being SM-like near 125 GeV.}
\vspace*{-6mm}
\label{table:nmp1nmp2nmp3}
\vspace{3mm}
\footnotesize
\begin{center}
\begin{tabular}{|l|r|r|r|}
\hline
{\bf Point} & NMP1 & NMP2 & NMP3
\\\hline\hline
$\tan \beta$  & 3 & 2 & 2
\\\hline
$\mu_{\rm eff}$ [GeV] & 200  & 200 & 200
\\\hline 
$\lambda$ & 0.64 & 0.6 & 0.57
\\\hline 
$\kappa$ & 0.25 & 0.18 & 0.2
\\\hline
$A_\lambda$ [GeV] & 560 & 405 & 395
\\\hline
$A_\kappa$ [GeV] & -10 & -10 & -80
\\\hline
%$m_{H_d}^2$  [GeV$^2$] & 371$^2$ 
%\\\hline
%$m_{H_u}^2$ [GeV$^2$] & 255$^2$ 
%\\\hline
%$m_{S}^2$ [GeV$^2$] & -107$^2$ 
%\\ \hline
$M_{\tilde{Q}_{3}} $ [GeV] & 650 & 700 & 530
\\ \hline  
$M_{\tilde{t}_R} $ [GeV] & 650 & 700 & 530
\\ \hline
$M_1$ [GeV] & 106 &  91 & 115
\\\hline
$M_2$ [GeV] & 200 & 200 & 200
\\\hline
$M_3$ [GeV] & 600 & 600 & 600
\\\hline\hline
\multicolumn{4}{|l|}{{\bf SM-like Higgs boson}}
\\\hline
$M_{H_1}$ [GeV] & 124.5 & 126.5 & 124.6
\\\hline
$R_{\gamma\gamma} (H_1)$  & 1.03 & 1.20 & 1.42
%$R_{\gamma\gamma} (H_1)$  & 1.079 (1.063) & 1.155 (1.239) & 1.221 (1.268)
\\\hline
$R_{VV} (H_1)$  & 0.94 & 1.02 & 1.12
%$R_{WW} (H_1)$  & 0.868 (0.8547) & 0.8698 (0.9332) & 0.882 (0.916)
\\\hline
$R_{b \bar b} (H_1) $ & 1.08 & 1.05 & 1.01
%$R_{b \bar b} (H_1) $ & 1.134 (1.117) & 1.011 (1.085) & 1.075 (1.115)
\\\hline
$R_{\Gamma_{tot}} (H_1)$ & 1.05 & 0.96 & 0.78
%& $4.072 \cdot 10^{-3}$ & $4.004 \cdot 10^{-3}$ & $3.055 \cdot 10^{-3}$ 
\\\hline
$R_{\sigma_{gg}} (H_1)$ & 0.97 & 0.96 & 0.77
\\\hline
$R_{incl} R_{\gamma \gamma} (H_1)$ & 1.00 & 1.15 & 1.11
\\\hline
$R_{incl} R_{VV} (H_1)$ & 0.91 & 0.98 & 0.88
\\\hline
$R_{\sigma_{tot}} (H_1)$& 0.92 & 0.99 & 0.89
\\\hline\hline
\multicolumn{4}{|l|}{{\bf Remaining Higgs spectrum} }
\\\hline
$M_{H_2}$ [GeV] & 155 & 132 & 129
\\\hline 
$M_{H_3}$ [GeV] & 637 & 465 & 456
\\\hline
$M_{A_1}$ [GeV] & 128 & 116 & 168
\\\hline
$M_{A_2}$ [GeV] & 634 & 463 & 454
\\\hline
$M_{H^\pm}$ [GeV] & 626 & 454 & 447
\\\hline\hline
\multicolumn{4}{|l|}{{\bf Sparticle masses and stop mixing}} 
\\ \hline
$m_{\tilde{g}}$  [GeV]  & 700 & 701 & 696
\\\hline
$m_{\tilde{\chi}^\pm_1}$ [GeV] & 137 & 131 & 131 \\ \hline
$m_{\tilde{\chi}^\pm_2}$ [GeV] & 281 & 284 & 284 \\ \hline
$m_{\tilde{\chi}^0_1}$ [GeV] & 78 & 68 & 85 \\ \hline 
$m_{\tilde{\chi}^0_2}$ [GeV] & 134 & 127 & 140 \\ \hline
$m_{\tilde{\chi}^0_3}$ [GeV] & 201 & 169 & 178\\ \hline
$m_{\tilde{\chi}^0_4}$ [GeV] & -231 & -232 & -227 \\ \hline
$m_{\tilde{\chi}^0_5}$ [GeV] & 292 & 290 & 290 \\ \hline
%$m_{\tilde{e}_{L,R}}=m_{\tilde{\mu}_{L,R}}$ [GeV] &1001  \\ \hline
%$m_{\tilde{\tau}_{1}}=m_{\tilde{\tau}_2}$ [GeV] & 1000 \\ \hline
%$m_{\tilde{\nu}_{e}}=m_{\tilde{\nu}_{\mu}}=m_{\tilde{\nu}_{\tau}}$ [GeV] & 999 \\ \hline
%$m_{\tilde{d}_{L.R}}=m_{\tilde{s}_{L.R}}$ [GeV] & 1021 \\ \hline
%$m_{\tilde{u}_{L,R}}=m_{\tilde{c}_{L.R}}$ [GeV] & 1019 \\ \hline
$m_{\tilde{b}_1}$ [GeV] & 667 & 715 & 538\\ \hline
$m_{\tilde{b}_2}$ [GeV] & 1014 & 1015 & 1011\\ \hline
$m_{\tilde{t}_1}$ [GeV] & 548 & 587 & 358\\ \hline
$m_{\tilde{t}_2}$ [GeV] & 782 & 838 & 686\\ \hline
$X_t/m_{\tilde{t}}$ & 1.74 & 1.86 & 2.26
\\\hline\hline
\multicolumn{4}{|l|}{{\bf Relic density}}
\\\hline
$\Omega h^2$ & 0.9819 & 0.1170 & 0.1100
\\ \hline
\end{tabular}\end{center}
\vspace*{-4mm}
\end{table}

%%%%%%%%%%%%%%%%%%%%%%%%%%%%%%%%%%%%%%%%%%%%%%%%%%%%%%%%%%%%%%%%%%

\noindent
\underline{\bf NMP2 with $(\lambda , \kappa) = (0.60,0.18)$} and $\tan \beta =2$ shows
a 126.5 GeV Higgs which has almost SM-like branching ratios into
massive vector bosons, $R_{VV}=1.02$. While it
will be difficult to distinguish this Higgs boson from the SM through
the decays into $WW,ZZ$, the branching ratio into $\gamma\gamma$ is
significantly larger than the SM value, $R_{\gamma
  \gamma}=1.20$, leading to an enhanced number of events in the photon
final state, $R_{\sigma_{incl}} R_{\gamma\gamma}=1.15$. The reason for the increased
$\gamma \gamma$ branching ratio is an enhanced photonic decay width
due to positive contributions from the SUSY loops
not present in the SM, in particular from the chargino loops, and a
reduced total width of $R_{\Gamma_{tot}}=0.96$. The slight deviation
of the branching ratio into $b\bar{b}$ from the SM value, $R_{b\overline{b}}=1.05$, is
within our theoretical error estimate.
The gluon fusion production cross-section ratio ends
up being close to the SM value for the same reasons as for NMP1, $R_{\sigma_{gg}}=0.96$.
Due to the SM-like decays into massive vector bosons the total Higgs
cross-section times branching ratios, which is mainly influenced by
the branching ratio into $WW$, is estimated to be very close to the SM
value, $R_{\sigma_{tot}}=0.99$. The heavier stop mass here
of 838 GeV is becoming more uncomfortable from the point of view of
fine-tuning. \sn

\noindent
\underline{\bf NMP3 with $(\lambda , \kappa) = (0.57,0.20)$} and $\tan \beta =2$ shows
a 124.6 GeV Higgs boson. Its distinguishing features are a large stop
mixing $X_t/m_{\tilde{t}}=2.26$ and a considerably enhanced branching
ratio (normalised to the SM) into photons,
$R_{\gamma\gamma}=1.42$. The large mixing helps to increase  
the Higgs boson mass as in Eq.~(\ref{eq:HiggsMassCorr}) and
consequently allows the stop masses  to be reduced, with the heavier stop mass of 686 GeV 
being apparently 
more acceptable from the point of view of fine-tuning, but at the price of an increased stop mixing angle. 
The coupling of the SM-like Higgs boson to down-type fermions is reduced by
about 15\% compared to the SM so that the total Higgs width dominated by the
decay into $b\bar{b}$ is reduced, $R_{\Gamma_{tot}}=0.78$ (normalised to the SM). 
Also the couplings to massive gauge bosons 
are reduced by 6\% implying smaller decay rates into $WW$,
$ZZ$. The branching ratios (normalised to the SM)
into $WW$, $ZZ$ are, however, enhanced, $R_{VV}$=1.12. This is due to the
reduced total width. The large enhancement of $R_{\gamma\gamma}$ is a combination of positive
contributions from the SUSY particle loops, mainly charginos, together
with a reduced total width.\footnote{The phenomenon of an enhanced
  decay width into photons due to a suppressed decay into $b\bar{b}$
  has also been discussed in \cite{photonh1} for the lighter Higgs
  $H_1$ and in \cite{Ellwanger:2011aa} for $H_2$ as well as in
  \cite{Hall:2011aa}.} While the coupling to the top quarks is
almost SM-like negative contributions 
from the stop quark loops with relatively light stops and a large
Higgs coupling to these states in this scenario decrease the gluon
fusion cross-section\footnote{Note that (s)bottom loops only play a
  minor role in all our scenarios as we have small values of $\tan\beta$.} so that 
it amounts only to about 77\% of the SM value, leading to only $\sim
10$\% more events in the $\gamma\gamma$ final state compared to the
SM, $R_{\sigma_{incl}} R_{\gamma\gamma}=1.11$, whereas the number of
$WW$ events is suppressed, $R_{\sigma_{incl}} R_{VV}=0.88$. 
Thanks to the enhanced branching ratios into $\gamma\gamma$, $ZZ$ and
$WW$, a total cross-section times branching ratios still compatible
with the LHC searches is achieved, $R_{\sigma_{tot}}=0.89$.

\subsection{\underline{NMSSM with Second Lightest Higgs being SM-like}}

Table~\ref{table:nmp4nmp5nmp6} shows three points NMP4, NMP5, NMP6
which satisfy NMSSM perturbativity up to the GUT scale
and lead to the SM-like Higgs boson always being the second lightest one $H_2$ near 125 GeV.
%The points are ordered in terms of increasing stop mixing $X_t$ scaled
%by $m_{\tilde{t}}$. 
The mass of the lightest Higgs boson $H_1$ is of
about 90-96 GeV. We have verified in each case that $H_1$ is not in conflict with LEP
limits due to its singlet character and hence strongly
reduced couplings to the SM particles. The gluino mass for these points ranges from 700 to 875 GeV\footnote{The higher value of
  the gluino mass for the benchmark point NMP6 results from the
  gaugino mass chosen to be $M_3=800$ GeV in order to fulfill
  approximate GUT relations between the soft SUSY breaking gaugino masses. This condition
  is not necessary, however, and choosing a lower value of $M_3$ and
  hence $m_{\tilde{g}}$ does not change our results in the Higgs
  sector.} and, as in all cases, the squark and slepton masses not shown are all about 1~TeV. \sn 
%%%%%%%%%%%%%%%%%%%%%%%%%%%%%%%%%%%%%%%%%%%%%%%%%%%%%%%%%%
\renewcommand{\arraystretch}{1.15}
\begin{table}[!h]
\caption{NMSSM benchmark points with the second lightest Higgs boson being SM-like near 125 GeV.}
\vspace*{-6mm}
\label{table:nmp4nmp5nmp6}
\vspace{3mm}
\footnotesize
\begin{center}
\begin{tabular}{|l|r|r|r|}
\hline
{\bf Point} & NMP4 & NMP5 & NMP6
\\\hline\hline
$\tan \beta$  & 3 & 3 & 2
\\\hline
$\mu_{\rm eff}$ [GeV] & 200  & 200 & 140
\\\hline 
$\lambda$ & 0.67  & 0.66 & 0.55
\\\hline 
$\kappa$ & 0.1 & 0.12 & 0.31
\\\hline
$A_\lambda$ [GeV] & 650 & 650 & 210
\\\hline
$A_\kappa$ [GeV] & -10 & -10 & -210
\\\hline
%$m_{H_d}^2$  [GeV$^2$] & 371$^2$ 
%\\\hline
%$m_{H_u}^2$ [GeV$^2$] & 255$^2$ 
%\\\hline
%$m_{S}^2$ [GeV$^2$] & -107$^2$ 
%\\ \hline
$M_{\tilde{Q}_{3}} $ [GeV] & 600 & 600 & 800
\\ \hline  
$M_{\tilde{t}_R} $ [GeV] & 600 & 600 & 600
\\ \hline
$M_1$ [GeV] & 200 & 200 & 145
\\\hline
$M_2$ [GeV] & 400 & 400 & 300
\\\hline
$M_3$ [GeV] & 600 & 600 & 800
\\\hline\hline
\multicolumn{4}{|l|}{{\bf SM-like Higgs boson}}
\\\hline
$M_{H_2}$ [GeV] & 123.8 & 126.5 & 124.5
\\\hline
$R_{\gamma\gamma} (H_2)$  & 1.06 & 1.15 & 1.39
%$R_{\gamma\gamma} (H_2)$  & 1.052 (1.094) & 1.069 (1.19) & 1.025 (1.38)
\\\hline
$R_{VV} (H_2)$  & 1.01 & 1.06 & 1.10
%$R_{VV} (H_2)$  & 0.876 (0.911) & 0.877 (0.977) & 0.756 (1.016)
\\\hline
$R_{b \bar b} (H_2) $ & 1.05 & 1.03 & 1.00
%$R_{b \bar b} (H_2) $ & 1.042 (1.083) & 0.951 (1.059) & 0.771 (1.037) 
\\ \hline
$R_{\Gamma_{tot}} (H_2)$  & 0.99 & 0.93 & 0.81
%& $3.772 \cdot 10^{-3}$ & $3.856 \cdot 10^{-3}$ &
%$3.141  \cdot 10^{-3}$
\\\hline
$R_{\sigma_{gg}} (H_2)$& 1.00 & 0.96 & 0.91
\\\hline
$R_{incl} R_{\gamma \gamma} (H_2)$ & 1.06 & 1.11 & 1.26
\\\hline
$R_{incl} R_{VV} (H_2)$ & 1.01 & 1.03 & 1.00
\\\hline
$R_{\sigma_{tot}} (H_2)$& 1.01 & 1.03 & 1.02
\\\hline\hline
\multicolumn{4}{|l|}{{\bf Remaining Higgs spectrum} }
\\\hline
$M_{H_1}$ [GeV] & 90 & 96 & 90
\\\hline 
$M_{H_3}$ [GeV] & 654 & 656 & 325
\\\hline
$M_{A_1}$ [GeV] & 86 & 93 & 249
\\\hline
$M_{A_2}$ [GeV] & 655 & 656 & 317
\\\hline
$M_{H^\pm}$ [GeV] & 643 & 645 & 312
\\\hline\hline
\multicolumn{4}{|l|}{{\bf Sparticle masses and stop mixing}} 
\\ \hline
$m_{\tilde{g}}$  [GeV] & 699 & 699 & 875
\\\hline
$m_{\tilde{\chi}^\pm_1}$ [GeV] & 182 & 182 & 114\\ \hline
$m_{\tilde{\chi}^\pm_2}$ [GeV] & 438 & 438 & 342\\ \hline
$m_{\tilde{\chi}^0_1}$ [GeV] & 69 & 78 & 80 \\ \hline 
$m_{\tilde{\chi}^0_2}$ [GeV] & 173 & 175 & 163\\ \hline
$m_{\tilde{\chi}^0_3}$ [GeV] & 233 & 238 & -169\\ \hline
$m_{\tilde{\chi}^0_4}$ [GeV] & -241 & -239 & 197\\ \hline
$m_{\tilde{\chi}^0_5}$ [GeV] & 438 & 438 & 343\\ \hline
$m_{\tilde{b}_1}$ [GeV] & 619 & 617 & 822\\ \hline
$m_{\tilde{b}_2}$ [GeV] & 1013 & 1013 & 827\\ \hline
$m_{\tilde{t}_1}$ [GeV] & 517 & 483 & 549\\ \hline
$m_{\tilde{t}_2}$ [GeV] & 724 & 741 & 892\\ \hline
$X_t/m_{\tilde{t}}$ & 1.56 & 1.89 & -1.83
\\\hline\hline
\multicolumn{4}{|l|}{{\bf Relic density}}
\\\hline
$\Omega h^2$ & 0.0999 & 0.1352 & 0.1258 
\\ \hline
\end{tabular}\end{center}
\vspace*{-6mm}
\end{table}
%%%%%%%%%%%%%%%%%%%%%%%%%%%%%%%%%%%%%%%%%%%%%%%%%%%%%%%%%%%%%%%%%%%

\noindent
\underline{\bf NMP4 with $(\lambda , \kappa) = (0.67,0.10)$} and  $\tan \beta =3$ shows
a 123.8 GeV Higgs boson with almost SM-like branching ratios into
massive vector bosons, $R_{VV}=1.01$. It differs from the SM Higgs
by its slightly larger value of branching ratio (normalised to the SM) into photons,
$R_{\gamma \gamma}=1.06$, which is due to the positive SUSY particle loop contributions,
mainly from chargino loops. The combined effect of the sbottom 
and stop loops in addition to the heavy quark loops is such that the
gluon fusion production cross-section is SM-like. The total production
cross-section times branching ratios is near the SM value,
$R_{\sigma_{tot}}=1.01$. With $X_t/m_{\tilde{t}}=1.56$, the heavier stop
mass here of 724 GeV is a little uncomfortable from the point of view
of fine-tuning, but not too bad. While this Higgs boson will be
difficult to be distinguished from the SM Higgs state, the main new interesting feature is
the appearance of two light bosons with $M_{H_1}=90$~GeV and
$M_{A_1}=86$ GeV which are predominantly composed 
of singlet $S$ and therefore would have escaped detection at
LEP. \sn

\noindent
\underline{\bf NMP5 with $(\lambda , \kappa) = (0.66,0.12)$} and  $\tan \beta =3$ shows
a 126.5 GeV Higgs boson which is also hard to distinguish from the SM
Higgs due to its gauge boson branching ratios being close to the SM
values, $R_{VV}=1.06$.
However the one into $\gamma\gamma$ is significantly larger,
$R_{\gamma \gamma }=1.15$, which is the combined effect of an 
enhancement due to SUSY particle loops and a
smaller total decay width, $R_{\Gamma_{tot}}= 0.93$, mainly due to a
smaller decay width into $b\bar{b}$ as consequence of a reduced Higgs
coupling to down-type fermions. With the additional suppression of
other decay channels, in the end the branching ratio into $b\bar{b}$
is slightly larger than in the SM, $R_{b\overline{b}}=1.03$. The
gluon fusion production cross-section is only slightly smaller than in
the SM, $R_{\sigma_{gg}}=0.96$. Together with the enhanced branching
ratios into gauge bosons, we estimate the total production Higgs cross-section 
times branching ratios to be close to the SM value,
$R_{\sigma_{tot}}=1.03$, while the number of events in the
$\gamma\gamma$ final state is enhanced compared to the SM,
$R_{\sigma_{incl}} R_{\gamma\gamma}=1.11$.
The heavier stop mass here of 741 GeV is a little uncomfortable from
the point of view of fine-tuning, but not too bad. There are two light bosons
with $M_{H_1}=96$ GeV and $M_{A_1}=93$ GeV which are predominantly composed
of singlet $S$ and therefore would have escaped detection at LEP. \sn

\noindent
\underline{\bf NMP6 with $(\lambda , \kappa) = (0.55,0.31)$} and  a
lower $\tan \beta =2$ shows a 124.5 GeV Higgs boson which can be
distinguished from the SM Higgs state by its branching ratios (normalised to
the SM) into massive gauge bosons of  $R_{VV}=1.10$ and in particular
by into photons. With $R_{\gamma \gamma }=1.39$ the $\gamma\gamma$ branching
ratio shows a very significant enhancement. Once again this is the combined
effect of a reduction of the total width, $R_{\Gamma_{tot}}=0.81$, and
a larger Higgs to photon coupling due to SUSY particle loops. The
former results from a reduced Higgs coupling to down-type quarks,
suppressing the partial width into $b\bar{b}$. Nevertheless, the
branching ratio for this final state is SM-like, $R_{b\overline{b}}=1.00$, due to the
additional suppression of other decay channels contributing to $\Gamma_{tot}$. 
In this scenario, the squark loops reduce the gluon fusion cross-section by about 9\%, $R_{\sigma_{gg}}=0.91$, resulting in an estimate
of the total cross-section times branching ratios which due to the
enhanced branching ratios into vector bosons is very close to the SM,
$R_{\sigma_{tot}}=1.02$. The number of photon final state events,
however, is enhanced by considerable 26\% compared to the SM. 
The mixing of $X_t/m_{\tilde{t}}=-1.83$ leads
to stop masses of 549 and 892 GeV, the latter making the scenario
more uncomfortable from the point of view of fine-tuning.
Interestingly, in this case 
there is only one light boson with $M_{H_1}=90$ GeV, which is predominantly
singlet-like and therefore would have escaped detection at LEP. 

%%%%%%%%%%%%%%%%%%%%%%%%%%%%%%%%%%%%%%%%%%%%%%%%%%%%%%%%%%%%%%%%%%
\subsection{\underline{NMSSM-with-extra-matter and Second Higgs being SM-like}}

Table~\ref{table:nmp7nmp8nmp9} shows three points NMP7, NMP8, NMP9
which satisfy perturbativity up to the GUT scale providing that the
NMSSM is supplemented by three $SO(10)$ $10$--plets 
as in Table~\ref{table:lambda3}. The only slightly larger value of $\lambda =0.68-0.69$ allows 
the extra matter to be at or above the TeV scale. It therefore may play no role
in any LHC phenomenology, apart from the indirect effect of allowing
$\lambda$ to be a bit larger than allowed by the usual requirement of
perturbativity in the NMSSM with no extra matter. All these points
lead to the SM-like Higgs boson always being the second lightest one $H_2$
near 125 GeV. The points are ordered in terms of increasing stop
mixing $X_t$ scaled by $m_{\tilde{t}}$. 
%%%%%%%%%%%%%%%%%%%%%%%%%%%%%%%%%%%%%%%%%%%%%%%%%%%%%%%%%%%%%%%%%%%
\renewcommand{\arraystretch}{1.15}
\begin{table}[!h]
\caption{NMSSM-with-extra-matter benchmark points with second lightest Higgs
  being SM-like near 125 GeV.} 
\vspace*{-6mm}
\label{table:nmp7nmp8nmp9}
\vspace{3mm}
\footnotesize
\begin{center}
\begin{tabular}{|l|r|r|r|}
\hline
{\bf Point} & NMP7 &  NMP8 & NMP9
\\\hline\hline
$\tan \beta$  & 2 & 2 &  3
\\\hline
$\mu_{\rm eff}$ [GeV] & 200 & 190 & 200 
\\\hline
$\lambda$ & 0.68  & 0.69 & 0.68
\\\hline 
$\kappa$ & 0.06 & 0.125 & 0.2
\\\hline
$A_\lambda$ [GeV] & 500 & 420 & 600  
\\\hline
$A_\kappa$ [GeV] & -100 & -100 & -10
\\\hline
%$m_{H_d}^2$  [GeV$^2$] & 570$^2$ & 562$^2$
%& 389$^2$ 
%\\\hline
%$m_{H_u}^2$ [GeV$^2$] & 192$^2$ & 166$^2$ & 185$^2$ 
%\\\hline
%$m_{S}^2$ [GeV$^2$] & 25$^2$ & 42$^2$ & 57$^2$
%\\ \hline
$M_{\tilde{Q}_{3}} $ [GeV] & 750 & 400 & 500
\\ \hline  
$M_{\tilde{t}_R} $ [GeV] & 750 & 230 & 500
\\ \hline  
$M_1$ (GeV) & 200 &  120&  135
\\\hline
$M_2$ (GeV) & 400& 335& 200
\\\hline
$M_3$ (GeV) & 800& 800& 600
\\\hline\hline
\multicolumn{4}{|l|}{{\bf SM-like Higgs boson}}
\\\hline
$M_{H_2}$ [GeV] & 124.5  & 126.2 & 125.8
\\\hline
$R_{\gamma\gamma} (H_2)$  & 1.17 & 0.87 & 1.78
\\\hline
$R_{VV} (H_2)$  & 1.06 & 0.87 & 1.41
\\\hline
$R_{b \bar b} (H_2) $ & 1.02 & 1.12 & 0.84
\\\hline
$R_{\Gamma_{tot}} (H_2)$ & 0.90 & 1.06 & 0.53
%& $3.510 \cdot 10^{-3}$ & $4.361 \cdot 10^{-3}$ & $2.166 \cdot 10^{-3}$
\\ \hline
$R_{\sigma_{gg}} (H_2)$ & 1.03 & 1.01 & 0.79
\\\hline
$R_{incl} R_{\gamma \gamma} (H_2)$ & 1.20 & 0.86 & 1.39
\\\hline
$R_{incl} R_{VV} (H_2)$ & 1.09 & 0.87 & 1.11
\\\hline
$R_{\sigma_{tot}} (H_2)$& 1.10 & 0.87 & 1.13
\\\hline\hline
\multicolumn{4}{|l|}{{\bf Remaining Higgs spectrum} }
\\\hline
$M_{H_1}$ [GeV] & 93 & 83  & 124
\\\hline
$M_{H_3}$ [GeV] & 495 & 434 & 641
\\\hline
$M_{A_1}$ [GeV] & 99 & 139 & 119
\\\hline
$M_{A_2}$ [GeV] & 503 & 438 & 640
\\\hline
$M_{H^\pm}$ [GeV] & 485 & 422 & 629
\\\hline\hline
\multicolumn{4}{|l|}{{\bf Sparticle masses and stop mixing}} 
\\ \hline
$m_{\tilde{g}}$  [GeV]  & 882 & 877 & 695
\\\hline
$m_{\tilde{\chi}^\pm_1}$ [GeV] & 179 & 159 & 137 \\ \hline
$m_{\tilde{\chi}^\pm_2}$ [GeV] & 439 & 379 & 281\\ \hline
$m_{\tilde{\chi}^0_1}$ [GeV] & 66 & 76 & 81  \\ \hline 
$m_{\tilde{\chi}^0_2}$ [GeV] & 160 & 120 & 143 \\ \hline
$m_{\tilde{\chi}^0_3}$ [GeV] & 231 & 191 & 185 \\ \hline
$m_{\tilde{\chi}^0_4}$ [GeV] & -248 & -233 & -237 \\ \hline
$m_{\tilde{\chi}^0_5}$ [GeV] & 440 & 380 & 291 \\ \hline
%$m_{\tilde{e}_{L,R}}=m_{\tilde{\mu}_{L,R}}$ [GeV] &1001  & 1001 & 1001 \\ \hline
%$m_{\tilde{\tau}_{1}}=m_{\tilde{\tau}_2}$ [GeV] & 1000 & 1000 & 1000 \\ \hline
%$m_{\tilde{\nu}_{e}}=m_{\tilde{\nu}_{\mu}}=m_{\tilde{\nu}_{\tau}}$ [GeV] & 998 & 998 & 998 \\ \hline
%$m_{\tilde{d}_{L.R}}=m_{\tilde{s}_{L.R}}$ [GeV] & 1021 & 1020 & 1029 \\ \hline
%$m_{\tilde{u}_{L,R}}=m_{\tilde{c}_{L.R}}$ [GeV] & 1020 & 1019 & 1028 \\ \hline
$m_{\tilde{b}_1}$ [GeV] & 779 & 379 & 514 \\ \hline
$m_{\tilde{b}_2}$ [GeV] & 1022 & 999 & 1011\\ \hline
$m_{\tilde{t}_1}$ [GeV] & 786 & 104 & 391\\ \hline
$m_{\tilde{t}_2}$ [GeV] & 797 & 442 & 634 \\ \hline
$X_t/m_{\tilde{t}}$ & 0 & 1.48  & 1.77 
\\\hline\hline
\multicolumn{4}{|l|}{{\bf Relic density}}
\\\hline
$\Omega h^2$ & 0.0973 & 0.1101 & 0.1031
\\ \hline
\end{tabular}\end{center}
\vspace*{-6mm}
\end{table}
%%%%%%%%%%%%%%%%%%%%%%%%%%%%%%%%%%%%%%%%%%%%%%%%%%%%%%%%%%%%%%%%%%%%
In two of the benchmark
scenarios $H_1$ is light with masses of 83 and 93 GeV, respectively, in one scenario it is
almost degenerate with $H_2$ with a mass of 124 GeV. In all these
cases we have verified that such light bosons 
are sufficiently singlet-like that they would have been produced with sufficiently low
rates in order not to be in conflict with LEP and LHC searches. The gluino mass ranges between
about 700 and 880 GeV. Recall that the squark and slepton
masses not shown are all about 1 TeV. \sn

\noindent
\underline{\bf NMP7 with $(\lambda , \kappa) = (0.68,0.06)$} and  $\tan \beta =2$ involves
a 124.5 GeV Higgs boson which will be difficult to be distinguished from the SM Higgs
by branching ratios into massive vector bosons (normalised to the SM)
of $R_{VV}=1.06$. The reduced coupling to down-type  
quarks leads to a smaller decay width into $b\bar{b}$ and hence a
smaller total width than in the SM, $R_{\Gamma_{tot}}=0.90$. 
The combined effect of a reduced decay into
$b\bar{b}$ and a positive contribution of, in particular, chargino loops
to the Higgs coupling to photons leads to a significantly enhanced
branching ratio into photons compared to the SM with
$R_{\gamma\gamma}=1.17$, which can be used to distinguish this NMSSM
scenario from the SM case through the enhanced number of events in the
photon final state, $R_{\sigma_{incl}} R_{\gamma\gamma}=1.20$. 
The $H_2$ coupling to up-type quarks is
SM-like so that the gluon fusion production dominated by (s)top quark
loops is not altered. With a small positive contribution from stop
loops this leads to a
3\% enhancement, $R_{\sigma_{gg}}=1.03$, and implies an estimated 
total Higgs production cross-section times branching ratio (normalised
to the SM value) of $R_{\sigma_{tot}}=1.10$.
The stop masses are almost 800 GeV making this point appear to be somewhat
fine-tuned. However, the absence of any stop mixing $X_t/m_{\tilde{t}}=0$, which is the
main feature of this point, serves to ameliorate the fine-tuning. 
Note that, in the MSSM, a 125 GeV Higgs with 
zero stop mixing would be impossible. In addition
there are two light bosons with $M_{H_1}=93$~GeV and $M_{A_1}=99$ GeV
which are predominantly singlet-like and therefore would have escaped detection at LEP.  \sn
 
\noindent
\underline{\bf NMP8 with $(\lambda , \kappa) = (0.69,0.125)$} and  $\tan \beta =2$ involves
a 126.2 GeV Higgs boson. The main features of this scenario are
reduced branching ratios not only into massive gauge bosons,
$R_{VV}=0.87$, but also into photons,
$R_{\gamma\gamma}=0.87$, and particularly light stop masses with the lighter
$m_{\tilde{t}_1}=104$ GeV. The branching ratio reductions are due to an enhanced
decay width into $b\bar{b}$ because of a slightly larger coupling to
down-type quarks than in the SM. In addition, the decay into photons
is reduced because of smaller $H_2$ couplings to up-type quarks and
negative stop quark loop contributions in this scenario. In the gluon
fusion production the positive contributions from the stop quark loops
counterbalance the reduction of the cross-section because of the smaller
$H_2$ coupling to top quarks, so that the ratio to the SM cross-section is nearly one, $R_{\sigma_{gg}}=1.01$. However the smaller
branching ratios into $WW,ZZ$ and $\gamma\gamma$ drive the total cross-section times branching ratios down to $R_{\sigma_{tot}}=0.87$
(normalised to the SM). This is still compatible with the present LHC
searches though, according to our criteria. With
$X_t/m_{\tilde{t}}=1.48$, the interesting feature of this scenario is
the very light stop which is close enough in mass to the lightest
neutralino in order to dominantly decay via the flavour changing
neutral current decay into charm and $\tilde{\chi}_1^0$. 
Such a light stop is not excluded in this decay channel by 
the present LHC searches. The correct amount of dark matter relic
density is here achieved
through $\tilde{t}_1 \tilde{\chi}_1^0$ co-annihilation into $W^+
b$. Also in this scenario we have two light bosons with $M_{H_1}=83$~GeV 
and $M_{A_1}=139$ GeV which are predominantly singlet-like and
hence not in conflict with LEP and LHC exclusion limits. \sn

\noindent
\underline{\bf NMP9 with $(\lambda , \kappa) = (0.68,0.20)$} and  $\tan \beta =3$ involves
a 125.8 GeV Higgs boson. The main features of this scenario are the
near degeneracy of the physical SM-like Higgs with a lighter Higgs boson of mass $M_{H_1}=124$
GeV as well as considerably enhanced branching ratios (normalised to
the SM) into photons, $R_{\gamma\gamma}=1.78$, and into massive gauge
bosons,  $R_{VV}=1.41$. Due to the near degeneracy of  
$H_1$ and $H_2$, and the associated large mixing,
each state contains a significant singlet component,
with the most SM-like Higgs $H_2$ coupling only with about 60\% of the SM strength
to down-type quarks so that both the decay width into $b\bar{b}$ and
the total width are reduced, $R_{b\bar{b}}=0.84$ and
$R_{\Gamma_{tot}}=0.53$,  and hence the branching ratios into gauge 
bosons enhanced. As the $H_2$ coupling to top quarks is also reduced
the gluon fusion production cross-section goes down supplemented by a
small negative contribution from stop loops, so that in the end we
have $R_{\sigma_{gg}}=0.79$. Nevertheless, due to the enhanced branching
ratios, the combined cross-section times branching ratios is compatible with LHC
searches as $R_{\sigma_{tot}}=1.13$. Particularly interesting is the
still considerably enhanced number of photon final state events compared
to the SM,  $R_{\sigma_{incl}} R_{\gamma\gamma}=1.39$, 
which is a distinctive feature of this scenario. Also the $H_1$ state is
consistent with LHC searches as its couplings to up-type quarks are reduced so that it
cannot be produced with sufficiently large rate to yield a second
signal at the LHC. In addition there is a  singlet-dominated boson of
mass $M_{A_1}=119$ GeV. With $X_t/M_{m_{\tilde{t}}}=1.77$, and 
the heavier stop mass here of 634 GeV, and a gluino mass of 695 GeV,
there is relatively little fine-tuning for this point.

%%%%%%%%%%%%%%%%%%%%%%%%%%%%%%%%%%%%%%%%%%%%%%%%%%%%%%%%%%%%%%%%%%%
\subsection{\underline{NMSSM-with-extra-matter and Lightest Higgs being SM-like}}

Table~\ref{table:nmp10nmp11nmp12} shows three points NMP10, NMP11, NMP12
which satisfy perturbativity up to the GUT scale providing that 
the NMSSM is supplemented by three $SO(10)$ $10$--plets 
as in Table~\ref{table:lambda2}. The larger value of $\lambda =0.7$,
held fixed for all these points, requires the extra matter to be close to the electroweak scale.
It  may therefore be expected to play a role in LHC phenomenology. 
All these points lead to the SM-like Higgs always being the lightest one, $H_1$, near 125 GeV.
The points are ordered in terms of increasing stop mixing $X_t$ scaled
by $m_{\tilde{t}}$. The second lightest Higgs boson has masses in the
range of about 130 to 145 GeV and, due to its singlet nature, we have checked that it is in
accordance with present LHC exclusion limits. The gluino mass for
these benchmark points is about 700  GeV and recall that the squark
and slepton masses not shown are all about 1 TeV. \sn

\noindent
\underline{\bf NMP10 with $(\lambda , \kappa) = (0.70,0.20)$} and  $\tan \beta =2$ involves
a 123.6 GeV Higgs which is SM-like in its decays into massive
vector bosons and $b$ quarks with the normalised branching ratios 
given by $R_{VV}=1.01$ and $R_{b\bar{b}}=1.04$. The
  branching ratio into photons, however, is enhanced with
  (normalised to the SM) $R_{\gamma \gamma}=1.15$. This is a
  consequence of positive contributions from chargino 
loops to the $H_1 \gamma\gamma$ coupling. The cross-section ratio 
$R_{\sigma_{gg}}=1.07$ is somewhat larger than in the SM because of
positive contributions from the stop loops. Together with the SM-like
branching ratios into massive gauge bosons and an enhanced one
into $\gamma\gamma$, in the end the estimated total Higgs cross-section times
branching ratios is not far from the SM value,
$R_{\sigma_{tot}}=1.08$. The number of photon events, however, is 
enhanced by significant 21\% compared to the SM. 
The stop masses are 630 and 645 GeV with the gluino mass
686 GeV, which does not require fine-tuning, since 
there is no stop mixing $X_t/m_{\tilde{t}}=0$.
This of course would be impossible in the MSSM.
The reason it is possible here is the large $\lambda = 0.7$, and low $\tan \beta =2$,
which together provide a nice tree-level Higgs mass contribution as in
Eq.~(\ref{eq:hmassNMSSM}). The heavier Higgs $H_2$ is not too far away
at 133 GeV but, of course, is singlet dominated and hence hard to
discover. \sn

%%%%%%%%%%%%%%%%%%%%%%%%%%%%%%%%%%%%%%%%%%%%%%%%%%%%%%%%%%%%%%%%%%%%
\renewcommand{\arraystretch}{1.15}
\begin{table}[!h]
\caption{NMSSM-with-extra-matter benchmark points with the lightest Higgs being SM-like near 125 GeV. }
\vspace*{-6mm}
\label{table:nmp10nmp11nmp12}
\vspace{3mm}
\footnotesize
\begin{center}
\begin{tabular}{|l|r|r|r|}
\hline
{\bf Point} & NMP10 &  NMP11 & NMP12
\\\hline\hline
$\tan \beta$  & 2 &  2 & 3
\\\hline
$\mu_{\rm eff}$ [GeV] & 200 & 200 & 200
\\\hline
$\lambda$  & 0.7  & 0.7  & 0.7
\\\hline 
$\kappa$  & 0.2 & 0.2 & 0.25
\\\hline
$A_\lambda$ [GeV]  & 405  & 405 & 560
\\\hline
$A_\kappa$ [GeV]  & -10  & -10  & -10
\\\hline
%$m_{H_d}^2$  [GeV$^2$]  & 363$^2$  & 354$^2$
%& 354$^2$ & 352$^2$  
%\\\hline
%$m_{H_u}^2$ [GeV$^2$]  & 159$^2$  & 177$^2$
%& 156$^2$ & 184$^2$
%\\\hline
%$m_{S}^2$ [GeV$^2$]  & -56$^2$   & -55$^2$ 
%& -54$^2$ & -55$^2$  
%\\ \hline
$M_{\tilde{Q}_{3}} $ [GeV] & 600 & 510 & 500
\\ \hline
$M_{\tilde{t}_R} $ [GeV] & 600 & 510 & 500
\\ \hline
$M_1$ [GeV] &120 & 120 & 125
\\\hline
$M_2$ [GeV] &200 &200 & 200
\\\hline
$M_3$ [GeV] &600 &600 & 600 
\\\hline\hline
\multicolumn{4}{|l|}{{\bf SM-like Higgs boson}}
\\\hline
$M_{H_1}$ [GeV] & 123.6 & 123.5 & 123.7
\\\hline
$R_{\gamma\gamma} (H_1)$  &  1.15 & 2.00 & 1.24
\\\hline
$R_{VV} (H_1)$  & 1.01 & 1.48 & 1.09
\\\hline
$R_{b \bar b} (H_1) $ & 1.04 & 0.82 & 1.01
\\\hline
$R_{\Gamma_{tot}} (H_1)$ & 0.98 & 0.40 & 0.90
%& $3.711 \cdot 10^{-03}$ & $1.519 \cdot 10^{-03}$
%& $3.422 \cdot 10^{-03}$
\\ \hline
$R_{\sigma_{gg}} (H_1)$ & 1.07 & 0.71 & 1.00
\\\hline
$R_{incl} R_{\gamma \gamma} (H_1)$ & 1.21 & 1.39 & 1.23
\\\hline
$R_{incl} R_{VV} (H_1)$ & 1.07 & 1.03 & 1.09
\\\hline
$R_{\sigma_{tot}} (H_1)$& 1.08 & 1.06 & 1.10
\\\hline\hline
\multicolumn{4}{|l|}{{\bf Remaining Higgs spectrum} }
\\\hline
$M_{H_2}$ [GeV] & 133 & 136 & 144
\\\hline
$M_{H_3}$ [GeV] & 468 & 458 & 629
\\\hline
$M_{A_1}$ [GeV] & 129 & 129 & 133
\\\hline
$M_{A_2}$ [GeV] & 469 & 459 & 626
\\\hline
$M_{H^\pm}$ [GeV] & 456 & 445 & 615
\\\hline\hline
\multicolumn{4}{|l|}{{\bf Sparticle masses and stop mixing}} 
\\ \hline
$m_{\tilde{g}}$  [GeV]  & 686 & 696 & 696
\\\hline
$m_{\tilde{\chi}^\pm_1}$ [GeV] & 131 & 131 & 137\\ \hline
$m_{\tilde{\chi}^\pm_2}$ [GeV] & 284 & 283 & 281\\ \hline
$m_{\tilde{\chi}^0_1}$ [GeV] & 82 & 82 & 83\\ \hline 
$m_{\tilde{\chi}^0_2}$ [GeV] & 137 & 137 & 141\\ \hline
$m_{\tilde{\chi}^0_3}$ [GeV] & 175 & 175 & 198\\ \hline
$m_{\tilde{\chi}^0_4}$ [GeV] & -241 & -240 & -236\\ \hline
$m_{\tilde{\chi}^0_5}$ [GeV] & 291 & 291& 293\\ \hline
%$m_{\tilde{e}_{L,R}}=m_{\tilde{\mu}_{L,R}}$ [GeV] & 1001 & 1001 & 1001 & 1001 \\ \hline
%$m_{\tilde{\tau}_{1}}=m_{\tilde{\tau}_2}$ [GeV] & 1000 & 1000 & 1000 &
%1000 \\ \hline
%$m_{\tilde{\nu}_{e}}=m_{\tilde{\nu}_{\mu}}=m_{\tilde{\nu}_{\tau}}$ [GeV] & 999 & 999 & 999 & 999 \\ \hline
%$m_{\tilde{d}_{L.R}}=m_{\tilde{s}_{L.R}}$ [GeV] & 1020 & 1020 & 1021 & 1021 \\ \hline
%$m_{\tilde{u}_{L,R}}=m_{\tilde{c}_{L.R}}$ [GeV] & 1019 & 1020 & 1020 & 1020\\ \hline
$m_{\tilde{b}_1}$ [GeV] & 622 & 525 & 516\\ \hline
$m_{\tilde{b}_2}$ [GeV] & 1013 & 1011 & 1011\\ \hline
$m_{\tilde{t}_1}$ [GeV] & 630 & 445 & 413\\ \hline
$m_{\tilde{t}_2}$ [GeV] & 645 & 620 & 624 \\ \hline
$X_t/m_{\tilde{t}}$ & 0 & 1.33 & 1.57
\\\hline\hline
\multicolumn{4}{|l|}{{\bf Relic density}}
\\\hline
$\Omega h^2$ & 0.1135 & 0.1161 & 0.1050
\\ \hline
\end{tabular}\end{center}
\vspace*{-4mm}
\end{table}
%%%%%%%%%%%%%%%%%%%%%%%%%%%%%%%%%%%%%%%%%%%%%%%%%%%%%%%%%%%%%%%%%%

\noindent
\underline{\bf NMP11 with $(\lambda , \kappa) = (0.70,0.20)$} and  $\tan \beta =2$ involves
a 123.5 GeV Higgs. Here we have a scenario with a strong singlet-doublet
mixing, this time for the lightest Higgs boson $H_1$. As a consequence,
its couplings to down-type quarks are reduced implying a smaller
decay width into $b\bar{b}$ leading to a total width of only about
40\% of the SM value, $R_{\Gamma_{tot}}=0.40$. The branching ratios
into massive gauge bosons are therefore significantly enhanced,
$R_{VV}=1.48$, despite $H_1$ also having reduced couplings to gauge
bosons. The most interesting feature, however, is the enhancement of the branching
ratio into photons by a factor 2, $R_{\gamma \gamma }=2.00$, due to the
combined effect of the smaller total width and the positive chargino loop
contributions which counterbalance the loss in the $H_1$ coupling
to $\gamma\gamma$ because of a reduced coupling to top quarks. 
The cross-section ratio $R_{\sigma_{gg}}=0.71$ results from the
reduced top loop contributions which cannot be made up for by positive
contributions from the squark loops. While the number of $WW$ final
states and the total cross-section times branching ratios, dominated
by the decay into $WW$, end up near their corresponding SM values, the number of
photonic events is enhanced by as much as 39\%, providing a distinctive
feature for this benchmark point. 
The nearby $H_2$ state with 136 GeV represents
no danger with respect to LHC exclusion limits as its couplings to SM
particles are reduced implying small enough production rates to be safe. 
With $X_t/m_{\tilde{t}}=1.33$, and a heavier stop mass of 620 GeV, and
a gluino mass of 696 GeV, this is a natural point without much fine-tuning.
\sn

\noindent
\underline{\bf NMP12 with $(\lambda , \kappa) = (0.70,0.25)$} and  $\tan \beta =3$ involves
a 123.7 GeV Higgs boson which is harder to distinguish from the SM Higgs by branching ratios
into massive vector bosons (normalised to the SM) of $R_{VV}=1.09$. However with $R_{\gamma
  \gamma }=1.24$ the branching ratio into photons is significantly
different from the SM and enhanced  once again because of positive
chargino loop contributions. As the $H_1$ coupling to top quarks is the same as
in the SM and the marginal contributions from squark loops add up to
almost zero, the gluon production cross-section is the same as in the
SM, $R_{\sigma_{gg}}=1.00$. With the enhanced branching ratios into
vector bosons the total production times branching
ratios ends up to be larger than in the SM, $R_{\sigma_{tot}}=1.10$,
and the number of photon final state events is enhanced by significant 23\%.
The normalized mixing $X_t/m_{\tilde{t}}=1.57$ features stop masses of
413 and 624~GeV. With the gluino mass of 696 GeV these are quite
natural values which does not require fine-tuning. 

%%%%%%%%%%%%%%%%%%%%%%%%%%%%%%%%%%%%%%%%%%%%%%%%%%%%%%%%%%%%%%%%%%
\section{Summary and Conclusion \label{summary}} 

The recent ATLAS and CMS indications of a SM-like Higgs boson near 125 GeV are consistent not only with the SM Higgs but also a SUSY Higgs. If it is a SUSY Higgs, the question is what sort of SUSY theory is responsible for it, and what part of parameter space of the SUSY theory does it correspond to?
In this paper we have been guided by both naturalness and minimality
to consider the possibility that the SM-like Higgs arises from the
NMSSM. Minimality alone would suggest the MSSM, however a 125 GeV
Higgs can only be achieved in the MSSM with stops necessarily having
some mixing and mass parameters above 1 TeV, which leads to some considerable
amount of tuning. By contrast, the NMSSM, which includes only one
additional singlet superfield $S$, allows a 125 GeV Higgs to arise
with all stop masses and mixing below 1~TeV.  Thus the NMSSM appears
to be the best compromise between naturalness and minimality that can
account for a 125 GeV SM-like Higgs boson, with the SM itself excluded on
naturalness grounds. \sn

Given that the NMSSM is a well-motivated SUSY theory, the next
question is what part of NMSSM parameter space does the SM-like Higgs
boson near 125 GeV correspond to? Clearly on naturalness grounds we
are led to consider large values of $\lambda$ and low values of $\tan
\beta$, in order to maximise the tree-level contribution to the Higgs
mass in Eq.~(\ref{eq:hmassNMSSM}). However $\lambda$ cannot be too large
otherwise it would blow up below the GUT scale which we regard as
unacceptable. Therefore it is important to know just how large
$\lambda$ can be, as a function of $\kappa$ and $\tan \beta$. 
We have studied this question using two-loop RGEs and found that it
cannot be as large as is often assumed in the literature. However, if
the SUSY desert contains extra matter, then $\lambda$ can be
increased, depending on the precise mass threshold of the extra
matter, which is a new result of this paper. \sn 

Assuming the NMSSM with large values of $\lambda$ and low values of
$\tan \beta$, which may or may not require
extra matter, another key requirement for us is having stop masses and
mixings below 1 TeV, on naturalness grounds. Since the LHC has already
placed strong limits on the first and second family squarks we are
therefore led to consider a general NMSSM, i.e. not the constrained
NMSSM. Given the rich parameter space of the NMSSM, it is practically 
impossible to perform scans over all NMSSM parameter space to identify
which regions are consistent with a 125 GeV Higgs. Instead we have
adopted a different strategy in this paper, namely that of using
benchmark points. Each benchmark point in the NMSSM parameter space
allows everything to be calculated to the standard of precision that
is currently available. For each benchmark point, the nature and
properties of the SM-like NMSSM Higgs boson can be scrutinised and compared
to the SM expectation, which could allow the NMSSM Higgs to be
resolved from the SM Higgs boson by future LHC searches and
measurements. It is possible for experimentalists to use these
benchmark points to test the NMSSM in different regions of the
parameter space by performing studies beyond what, as mere
theorists, we are capable of.\sn

Motivated by these considerations we have proposed four sets of
benchmark points corresponding to the SM-like Higgs being the lightest
or the second lightest in the NMSSM or the NMSSM-with-extra-matter. 
Points NMP1, NMP2, NMP3 are for the 
NMSSM with the lightest Higgs being SM-like near 125 GeV.
Points NMP4, NMP5, NMP6 are for the 
NMSSM with the second lightest Higgs being SM-like near 125 GeV.
Points NMP7, NMP8, NMP9 are for the 
NMSSM-with-extra-matter at or above the TeV scale where 
the second lightest Higgs is SM-like near 125 GeV.
Points NMP10, NMP11, NMP12 are for the 
NMSSM-with-extra-matter close to the electroweak scale and 
the lightest Higgs being SM-like near 125 GeV.
For each of these points we have 
discussed how the NMSSM Higgs boson near 125 GeV may be distinguished
from the SM Higgs state in future LHC searches. 
We have considered both the gluon gluon cross-section and the branching ratios 
into WW, ZZ and two photons. We also considered a weighted total Higgs
cross-section times branching ratios as a measure of how many Higgs
events would be observed in current searches. \sn

Each benchmark point, with different values of Higgs cross-sections
and branching ratios, acts like a little beacon which in principle
can allow experiment to guide us
to the correct part of the NMSSM parameter space. It is worth emphasising,
however, that the values of $R_{VV}$ and $R_{b\bar{b}}$ quoted here are only
accurate to a few percent, as they are obtained by using two different
programs for the calculation of the branching ratios, {\tt NMSSMTools}
for the NMSSM and {\tt HDECAY} for the SM case, which use different
approximations in the calculation of the various partial widths.
The differences are always less than or equal to 5\% for the quoted
values, while the maximum deviations for $R_{\gamma \gamma}$ differ by up to
10\% . Although these theoretical
uncertainties are well within the expected experimental error for the measurement
of these branching ratios at the LHC, it does mean that resolving
the NMSSM Higgs from the SM Higgs is going to be challenging.
However, in some cases the Higgs branching ratio into two photons may easily be
enhanced by a factor of two in the NMSSM, as seen in NMP9 and
NMP11. These very large enhancements result in part from the  
reduction in the bottom quark-antiquark decay rate due to the reduced
amount of $H_d$ component in the SM-like Higgs boson. Moreover
significant enhancements in the two photon channel are also seen for
other points where the $b\bar{b}$ decay rate is not
reduced, and this is typically due to the effect of chargino loops. \sn

In general we have taken the stop masses and mixing to be as low as
possible, the extreme example of this being NMP8 where the lightest
stop is only 104 GeV. We find it remarkable that such light stops 
are still allowed by current searches, although we would expect stops below 1 TeV,
as assumed here, to be discovered soon at the LHC. 
We look forward to the forthcoming LHC run in 2012 with much excitement.

%%%%%%%%%%%%%%%%%%%%%%%%%%%%%%%%%%%%%%%%%%%%%%%%%%%%%%%%%%%%%%%%%
\subsubsection*{Acknowledgments} 
SFK would like to thank S. Belyaev and H. Haber for stimulating
discussions. MMM gratefully acknowledges several discussions with
G. B\'elanger, J. Cao, U. Ellwanger, A. Pukhov and M. Spira. R.~Nevzorov would
like to thank X. Tata for fruitful discussions.
MMM is supported by the DFG SFB/TR9  ``Computational Particle Physics''. 
The work of RN was supported by the U.S. Department of Energy under 
Contract DE-FG02-04ER41291. SFK acknowledges partial support 
from the STFC Consolidated ST/J000396/1 and EU ITN grant UNILHC 237920 (Unification in
the LHC era).

%%%%%%%%%%%%%%%%%%%%%%%%%%%%%%%%%%%%%%%%%%%%%%%%%%%%%%%%%%%%%%%%%

\newpage
\section*{Appendix}
\begin{appendix}

\section{\label{sec} Renormalisation Group Equations to Second Order}
\cleqn

The running of the gauge couplings from the GUT scale to the EW scale 
is determined by a set of RGEs. In our analysis, we use two--loop RGEs 
for the gauge and Yukawa couplings. Retaining only $\lambda$, $\kappa$,
$h_t$, $h_b$ and $h_{\tau}$ the two--loop RGEs in the NMSSM can be written as
\be
\ba{rcl}
\ds\frac{d g_i}{dt}&=&\ds\frac{\beta_i g_i^3}{(4\pi)^2}\,,\\[6mm]
\ds\frac{d\lambda}{dt}&=&\ds\frac{\lambda}{(4\pi)^2}\biggl[4\lambda^2 + 2\kappa^2 +
3h_t^2+3h_b^2+h_{\tau}^2-3g_2^2-\dfrac{3}{5}g_1^2+\dfrac{\beta^{(2)}_{\lambda}}{(4\pi)^2}\biggr]\,,\\[6mm]
\ds\frac{d\kappa}{dt}&=&\ds\frac{\kappa}{(4\pi)^2}\biggl[6 (\kappa^2 + \lambda^2)
+\frac{1}{(4\pi)^2} \biggl( 18 \lambda^2 g_2^2 + \ds\frac{18}{5} \lambda^2 g_1^2 \\[6mm] 
&-&\lambda^2 ( 12 \lambda^2 + 24 \kappa^2 + 18 h_t^2 + 18 h_b^2 + 6 h_{\tau}^2) - 24 \kappa^4 \biggr)\biggr]\,,\\[4mm]
\ds\frac{dh_t}{dt}&=&\ds\frac{h_t}{(4\pi)^2}\biggl[\lambda^2+6h_t^2+h_b^2-\ds\frac{16}{3}g_3^2
-3g_2^2-\ds\frac{13}{15}g_1^2+\frac{\beta^{(2)}_{h_t}}{(4\pi)^2}\biggr]\,,\\[6mm]
\ds\frac{dh_b}{dt}&=&\ds\frac{h_b}{(4\pi)^2}\biggl[\lambda^2+h_t^2+6h_b^2+h_{\tau}^2
-\ds\frac{16}{3}g_3^2-3g_2^2-\ds\frac{7}{15}g_1^2+\frac{\beta^{(2)}_{h_b}}{(4\pi)^2}\biggr]\,,\\[6mm]
\ds\frac{dh_{\tau}}{dt}&=&\ds\frac{h_{\tau}}{(4\pi)^2}\biggl[\lambda^2+3h_b^2+4h_{\tau}^2-3g_2^2-\frac{9}{5}g_1^2
+\frac{\beta^{(2)}_{h_{\tau}}}{(4\pi)^2}
\biggr]\,,
\ea
\label{eq4}
\ee
where $t=\ln\left[Q/M_X\right]$; the index $i$ runs from $1$ to $3$
and is associated with $U(1)_Y$, $SU(2)_W$ and $SU(3)_C$ gauge interactions; 
$\beta^{(2)}_{\lambda}$, $\beta^{(2)}_{h_t}$, $\beta^{(2)}_{h_b}$ and 
$\beta^{(2)}_{h_{\tau}}$ are the two--loop contributions to the 
corresponding $\beta$--functions. \sn

In the NMSSM the two--loop $\beta$--functions of the gauge couplings, 
$\beta^{(2)}_{\lambda}$, $\beta^{(2)}_{h_t}$, $\beta^{(2)}_{h_b}$ and
$\beta^{(2)}_{h_{\tau}}$ are given by\footnote{Note that the two-loop
  RGE results differ somewhat from the original two-loop 
RGE results first obtained in the NMSSM by King and White \cite{genNMSSM2},
which is why we have written them out explicitly here.
Of course the two-loop RGE results including extra matter are new
results presented in this paper for the first time.}
\be
\ba{rcl}
\beta_3&=&-3 + \ds\frac{1}{16\pi^2}\Biggl[ 14 g_3^2 + 9 g_2^2 + \ds\frac{11}{5} g_1^2
- 4 h_t^2 - 4 h_b^2 \Biggr]\,,\\[6mm]
\beta_2&=& 1 + \ds\frac{1}{16\pi^2}\Biggl[ 24 g_3^2 + 25 g_2^2+ \ds\frac{9}{5} g_1^2
- 6 h_t^2 - 6 h_b^2 - 2 h_{\tau}^2 - 2 \lambda^2 \Biggr]\,,\\[6mm]
\beta_1&=& \ds\frac{33}{5} + \ds\frac{1}{16\pi^2}\Biggl[\ds\frac{88}{5} g_3^2 + \ds\frac{27}{5} g_2^2+
\ds\frac{199}{25} g_1^2 - \ds\frac{26}{5} h_t^2 - \ds\frac{14}{5}h_b^2 -
\ds\frac{18}{5} h_{\tau}^2 - \ds\frac{6}{5} \lambda^2 \Biggr]\,,
\ea
\label{eq3}
\ee
\be
\ba{rcl}
\beta^{(2)}_{\lambda} & = & - 9 h_t^4 - 9 h_b^4 - 6 h_t^2 h_b^2 - 3 h_{\tau}^4 - 8 \kappa^4
- \lambda^2 \biggl( 9 h_t^2 + 9 h_b^2 + 3 h_{\tau}^2 + 12 \kappa^2 + 10 \lambda^2 \biggr) \\[3mm]
& + & 16 g_3^2 \biggl(h_t^2+h_b^2\biggr) + 6 g_2^2 \lambda^2
+ g_1^2 \left(\ds\frac{4}{5} h_t^2 - \ds\frac{2}{5} h_b^2 + \ds\frac{6}{5}h_{\tau}^2 
+ \ds\frac{6}{5} \lambda^2 \right)\\[3mm]
& + & \ds\frac{15}{2} g_2^4 + \ds\frac{9}{5} g_2^2 g_1^2 + \ds\frac{207}{50} g_1^4\,,\\[3mm]
\beta^{(2)}_{h_t}&=&-22 h_t^4 - 5 h_b^4 - 5 h_t^2 h_b^2 - h_b^2 h_{\tau}^2
-\lambda^2\biggl(3 \lambda^2 + 3 h_t^2 + 4 h_b^2 + h_{\tau}^2 + 2\kappa^2 \biggr)\\[3mm]
& + & 16 g_3^2 h_t^2 + 6 g_2^2 h_t^2 + g_1^2 \left(\ds\frac{6}{5}h_t^2+\ds\frac{2}{5}h_b^2\right)
- \ds\frac{16}{9}g_3^4 + \ds\frac{15}{2} g_2^4 + \ds\frac{2743}{450} g_1^4\\[3mm]
& + & 8 g_3^2 g_2^2 + \ds\frac{136}{45} g_3^2 g_1^2 + g_2^2 g_1^2\,,\\[3mm]
\beta^{(2)}_{h_b}&=& - 5 h_t^4 - 22 h_b^4 - 5 h_t^2 h_b^2 - 3 h_b^2 h_{\tau}^2 - 3 h_{\tau}^4
-\lambda^2\biggl(3 \lambda^2 + 4 h_t^2 + 3 h_b^2 + 2\kappa^2 \biggr) \\[3mm]
& + & 16 g_3^2 h_b^2 + 6 g_2^2 h_b^2 +
g_1^2\left(\ds\frac{4}{5} h_t^2 + \ds\frac{2}{5} h_b^2 + \ds\frac{6}{5} h_{\tau}^2\right)
- \ds\frac{16}{9} g_3^4 + \ds\frac{15}{2} g_2^4 + \ds\frac{287}{90} g_1^4\\[3mm]
& + & 8 g_3^2 g_2^2 + \ds\frac{8}{9} g_3^2 g_1^2 + g_2^2 g_1^2\,,\\[3mm]
\beta^{(2)}_{h_{\tau}}&=&-9 h_b^4 - 3 h_t^2 h_b^2 - 9 h_b^2 h_{\tau}^2
-10 h_{\tau}^4 - \lambda^2 \biggl(3 \lambda^2 + 3 h_t^2 + 3 h_{\tau}^2 
+ 2 \kappa^2 \biggr)\\[3mm] 
&+ & 16 g_3^2 h_b^2 + 6 g_2^2 h_{\tau}^2+
g_1^2 \left(-\ds\frac{2}{5}h_b^2+\ds\frac{6}{5}h_{\tau}^2\right)
+\ds\frac{15}{2} g_2^4 + \ds\frac{27}{2} g_1^4 + \ds\frac{9}{5}g_2^2g_1^2\,.
\ea
\ee

In the NMSSM with three extra $SO(10)$ 10-plets the two--loop $\beta$--functions of the 
gauge couplings, $\beta^{(2)}_{\lambda}$, $\beta^{(2)}_{h_t}$, $\beta^{(2)}_{h_b}$ and
$\beta^{(2)}_{h_{\tau}}$ can be written as
\be
\ba{rcl}
\beta_3&=&\ds\frac{1}{16\pi^2}\Biggl[ 48 g_3^2 + 9 g_2^2 + 3 g_1^2
- 4 h_t^2 - 4 h_b^2 \Biggr]\,,\\[3mm]
\beta_2&=& 4 + \ds\frac{1}{16\pi^2}\Biggl[ 24 g_3^2 + 46 g_2^2+ \ds\frac{18}{5} g_1^2
- 6 h_t^2 - 6 h_b^2 - 2 h_{\tau}^2 - 2 \lambda^2 \Biggr]\,,\\[3mm]
\beta_1&=& \ds\frac{48}{5} + \ds\frac{1}{16\pi^2}\Biggl[24 g_3^2 + \ds\frac{54}{5} g_2^2+
\ds\frac{234}{25} g_1^2 - \ds\frac{26}{5} h_t^2 - \ds\frac{14}{5}h_b^2 - 
\ds\frac{18}{5} h_{\tau}^2 - \ds\frac{6}{5} \lambda^2 \Biggr]\,,\qquad\qquad\\
\ea
\ee
\be
\ba{rcl}
\beta^{(2)}_{\lambda} & = & - 9 h_t^4 - 9 h_b^4 - 6 h_t^2 h_b^2 - 3 h_{\tau}^4 - 8 \kappa^4
- \lambda^2 \biggl( 9 h_t^2 + 9 h_b^2 + 3 h_{\tau}^2 + 12 \kappa^2 + 10 \lambda^2 \biggr) \\[3mm]
& + & 16 g_3^2 \biggl(h_t^2+h_b^2\biggr) + 6 g_2^2 \lambda^2
+ g_1^2 \left(\ds\frac{4}{5} h_t^2 - \ds\frac{2}{5} h_b^2 + \ds\frac{6}{5}h_{\tau}^2 + \ds\frac{6}{5} \lambda^2 \right)
\\[3mm]
& + & \ds\frac{33}{2} g_2^4 + \ds\frac{9}{5} g_2^2 g_1^2 + \ds\frac{297}{50} g_1^4\,,\\[3mm]
\beta^{(2)}_{h_t}&=&-22 h_t^4 - 5 h_b^4 - 5 h_t^2 h_b^2 - h_b^2 h_{\tau}^2
-\lambda^2\biggl(3 \lambda^2 + 3 h_t^2 + 4 h_b^2 + h_{\tau}^2 + 2\kappa^2 \biggr)\\[3mm]
& + & 16 g_3^2 h_t^2 + 6 g_2^2 h_t^2 + g_1^2 \left(\ds\frac{6}{5}h_t^2+\ds\frac{2}{5}h_b^2\right)
+ \ds\frac{128}{9}g_3^4 + \ds\frac{33}{2} g_2^4 + \ds\frac{3913}{450} g_1^4\\[3mm]
& + & 8 g_3^2 g_2^2 + \ds\frac{136}{45} g_3^2 g_1^2 + g_2^2 g_1^2\,,
\ea
\ee
$$
\ba{rcl}
\beta^{(2)}_{h_b}&=& - 5 h_t^4 - 22 h_b^4 - 5 h_t^2 h_b^2 - 3 h_b^2 h_{\tau}^2 - 3 h_{\tau}^4
-\lambda^2\biggl(3 \lambda^2 + 4 h_t^2 + 3 h_b^2 + 2\kappa^2 \biggr) \\[3mm]
& + & 16 g_3^2 h_b^2 + 6 g_2^2 h_b^2 +
g_1^2\left(\ds\frac{4}{5} h_t^2 + \ds\frac{2}{5} h_b^2 + \ds\frac{6}{5} h_{\tau}^2\right)
+ \ds\frac{128}{9} g_3^4 + \ds\frac{33}{2} g_2^4 + \ds\frac{413}{90} g_1^4\\[3mm]
& + & 8 g_3^2 g_2^2 + \ds\frac{8}{9} g_3^2 g_1^2 + g_2^2 g_1^2\,,\\[3mm]
\beta^{(2)}_{h_{\tau}}&=&-9 h_b^4 - 3 h_t^2 h_b^2 - 9 h_b^2 h_{\tau}^2
-10 h_{\tau}^4 - \lambda^2 \biggl(3 \lambda^2 + 3 h_t^2 + 3 h_{\tau}^2
+ 2 \kappa^2 \biggr)\\[3mm]
&+ & 16 g_3^2 h_b^2 + 6 g_2^2 h_{\tau}^2+
g_1^2 \left(-\ds\frac{2}{5}h_b^2+\ds\frac{6}{5}h_{\tau}^2\right)
+\ds\frac{33}{2} g_2^4 + \ds\frac{189}{10} g_1^4 + \ds\frac{9}{5}g_2^2g_1^2\,.
\ea
$$

\end{appendix}
%%%%%%%%%%%%%%%%%%%%%%%%%%%%%%%%%%%%%%%%%%%%%%%%%%%%%%%%%%%%%%%%%

\end{document}